\documentclass[aps,preprint,showpacs,preprintnumbers,amsmath,amssymb]{revtex4}
\usepackage{amsmath,mathrsfs,amsbsy,color,graphicx,bm,amsthm,amsfonts}
\usepackage{units}
\usepackage{bbm}
\usepackage{times}
\usepackage{dcolumn}
\usepackage{mathrsfs}
\usepackage{amsmath,amssymb,epsfig}
\usepackage{amsmath}
\usepackage{subfig}
\newcommand{\udots}{\mathinner{\mskip1mu\raise1pt\vbox{\kern7pt\hbox{.}}
\mskip2mu\raise4pt\hbox{.}\mskip2mu\raise7pt\hbox{.}\mskip1mu}}
\begin{document}
\title{Does fermionic entanglement always outperform bosonic entanglement in dilaton black hole?}
\author{ Wen-Mei Li$^1$, Jianbo Lu$^1$\footnote{lvjianbo819@163.com  (corresponding author)}, Shu-Min Wu$^1$\footnote{smwu@lnnu.edu.cn (corresponding author)} }
\affiliation{$^1$ Department of Physics, Liaoning Normal University, Dalian 116029, China}


\begin{abstract}
It has traditionally been believed that fermionic entanglement generally outperforms bosonic entanglement in relativistic frameworks, and that bosonic entanglement experiences sudden death in extreme gravitational environments. In this study, we analyze the genuine N-partite entanglement, measured by negativity, of bosonic and fermionic GHZ states, focusing on scenarios where a subset of  $m$ ($m<N$) constituents interacts with Hawking radiation generated by a Garfinkle-Horowitz-Strominger (GHS) dilaton black hole. Surprisingly, we find that quantum entanglement between the non-gravitational and gravitational modes for the bosonic field is stronger than that in the same modes for the fermionic field within dilaton spacetime.
This study challenges the traditional belief that  ``fermionic  entanglement always outperforms bosonic entanglement" in the relativistic framework. However, quantum entanglement between the gravitational modes and the combined gravitational and non-gravitational modes is weaker for the bosonic field than for the fermionic field in the presence of a dilaton black hole. 
Finally, the connection between the global  N-partite entanglement in the bosonic field and that in the fermionic field is influenced by the gravitational field's intensity. Our study reveals the intrinsic relationship between quantum entanglement of bosonic and fermionic fields in curved spacetime from a new perspective, and provides theoretical guidance for selecting appropriate field-based quantum resources for relativistic quantum information tasks under extreme gravitational conditions.
\end{abstract}

\vspace*{0.5cm}
\maketitle

\section{Introduction}
Quantum entanglement is an essential characteristic of quantum mechanics and a central resource in quantum information theory, as it enables nonclassical correlations between spatially separated subsystems. A central goal of quantum communication is the construction of large-scale quantum networks, raising key questions about state generation, required resources, and efficient verification \cite{LW1,LW2,LW3,LW4,LW5,LW6,LW7,LW8,LW9,LW10,LW11,LW12}. In this context, the generation of genuinely multipartite entangled states has attracted considerable attention in both theoretical \cite{LW13,LW14,LW15} and experimental studies \cite{LW16,LW17,LW18,LW19,LW20}, as such states are indispensable for fully exploiting the capabilities of quantum networks. Genuine multipartite entangled states are fully inseparable across all possible bipartitions and cannot be expressed as statistical mixtures of states that are separable under different partitions, commonly referred to as biseparable states \cite{LW21}. Multipartite entanglement plays a pivotal role in a wide range of quantum information tasks, including quantum metrology \cite{LW22}, quantum computation—such as measurement-based quantum computation \cite{LW23,LW24} and quantum error correction \cite{LW25} as well as quantum communication protocols like quantum key distribution \cite{LW26,LW27}, conference key agreement \cite{LW28}, and communication within quantum networks \cite{LW29}. In certain applications, genuine multipartite entanglement is an essential and irreplaceable resource \cite{LW26,LW30}.

Relativistic quantum information science is a rapidly advancing interdisciplinary field that integrates the core principles of quantum information theory, quantum field theory, and general relativity. Significant progress has been made in both theoretical and experimental aspects of this field in recent years \cite{SDF1,SDF2,SDF3,SDF4,SDF5,SDF6,SDF7,SDF8,SDF9,SDF10,SDF11,SDF12,SDF13,SDF14,SDF15,SDF16,SDF17,SDF18,SDF19,SDF20,SDF21,SDF22,SDF23,SDF24,SDF25,SDF26,SDF27,SDF28,SDF29,SDF30,SDF31,SDF32,SDF33,SDF34,SDF35,SDF36,SDF37,SDF38,SDF39,SDF40,SDF41,SDF42,SDF43,SDF44,SDF45,SDF46,SDF47,SDF48,SDF49,SDF50,SDF51,SDF52,
SDF53,SDF54,SDF55,SDF56,SDF57,AGL1,AGL2,AGL3,AGL99,AGL100,AGL101,AGL102,AGL103,AGL104}. Theoretically, extensive studies have been conducted on the effects of the Unruh effect in Rindler spacetime and the Hawking effect in black holes on quantum correlations and coherence, uncovering the mechanisms behind the generation and decoherence of these quantum resources in curved spacetime. It is widely believed that in non-inertial reference frames, for a bipartite system, the quantum entanglement of the bosonic field diminishes with increasing acceleration, eventually vanishing at infinite acceleration \cite{SDF7}. In contrast, the quantum entanglement of the fermionic field remains greater than that of the bosonic field, persisting even at infinite acceleration \cite{SDF8}. In the background of black hole spacetimes, bipartite fermionic entanglement continues to exceed that of bosonic entanglement \cite{SDF1,SDF2,SDF3,SDF4,SDF5,SDF6}. This leads to the traditional academic viewpoint that, within the relativistic framework, fermionic quantum correlations are regarded as more robust than bosonic correlations \cite{SDF1,SDF2,SDF3,SDF4,SDF5,SDF6,SDF7,SDF8,SDF9,SDF10,SDF11,SDF12}, making fermionic entanglement a more valuable quantum resource. Motivated by these observations, we seek to investigate the genuine N-partite entanglement of both bosonic and fermionic fields in the black hole backgrounds, aiming to test whether the traditional view that fermionic entanglement always outperforms bosonic entanglement holds when extended to multipartite systems. This investigation forms one of the primary motivations for our study. As quantum information tasks grow more intricate, the need for multipartite entanglement to effectively address relativistic quantum information challenges becomes ever more pressing. Thus, another key motivation of our research is to explore the properties of genuine N-partite entanglement in gravitational backgrounds and to identify the optimal quantum resources required for handling such tasks in relativistic settings.

This paper explores the genuine N-partite entanglement, quantified through negativity, of bosonic and fermionic GHZ states configurations in the presence of a GHS dilaton black hole. We begin by considering a scenario where 
$N$ observers share N-partite GHZ states in the asymptotically flat region. Subsequently, a gravitational environment is introduced in which $N-m$ observers remain inertial in flat spacetime, while 
$m$ ($m<N$) observers are positioned near the event horizon of the dilaton black hole. The GHS dilaton black hole, a compact object arising from superstring theory, is widely regarded as one of the most promising candidates for bridging quantum mechanics and gravity \cite{J9,J10,J11}. Our study reveals intriguing results: quantum entanglement between the non-gravitational and gravitational modes of the bosonic field is found to be stronger than in the fermionic field within dilaton spacetime. However, when considering the quantum entanglement between the gravitational modes and the combined gravitational and non-gravitational modes, the bosonic field exhibits weaker entanglement than the fermionic field under the influence of the dilaton black hole. Notably,  with increase of the dilaton, the global N-partite entanglement in the bosonic field becomes slightly smaller than that in the fermionic field, but after further changes, the global N-partite entanglement in the bosonic field becomes larger than that in the fermionic field. Therefore, the connection between the global  N-partite entanglement in the bosonic field and that in the fermionic field is influenced by the gravitational field's intensity. These findings challenge the traditional view that fermionic entanglement universally outperforms bosonic entanglement \cite{SDF1,SDF2,SDF3,SDF4,SDF5,SDF6,SDF7,SDF8}, particularly in the context of black hole spacetimes. Our work not only contributes to the ongoing debate on the relative robustness of fermionic versus bosonic entanglement but also offers novel insights into the behavior of quantum resources in extreme gravitational settings.

The paper is organized as follows.  Sec. II presents the quantization of bosonic and fermionic fields in the background of a GHS dilaton black hole. Sec. III introduces the concepts of one-tangle, two-tangle, and $\pi$-tangle. In Sec. IV, we explore the N-partite genuine entanglement of GHZ states for both bosonic and fermionic fields in the GHS dilaton spacetime. The final section provides a brief conclusion.

\section{Quantization of bosonic and fermionic fields in GHS dilaton spacetime \label{GSCDGE}}
The metric for a GHS dilaton black hole is given by 
\begin{eqnarray}\label{S1}
ds^{2}=-\bigg(\frac{r-2M}{r-2\epsilon}\bigg)dt^{2}+\bigg(\frac{r-2M}{r-2\epsilon}\bigg)^{-1}dr^{2}+r(r-2\epsilon)d\Omega^{2},
\end{eqnarray}
where $\epsilon$ and $M$ are parameters associated with the dilaton field and the black hole mass, respectively \cite{A1}. They are related by $\epsilon=Q^{2}/2M$, with $Q$ denoting the electric charge. In this paper, we consider $G=c=\hbar=\kappa_{B}=1$.

\subsection{Bosonic field}
The couple massive scalar field of general perturbation equation in this dilaton spacetime obeys the Klein-Gordon equation 
\begin{eqnarray}\label{S2}
\frac{1}{\sqrt{-g}}\partial_{\mu}(\sqrt{-g}g^{\mu\nu}\partial_{\nu})\Psi-(\mu+\xi R)\Psi=0,
\end{eqnarray}
where $\mu$ is the mass of the particle, $R$ is the Ricci scalar curvature, and $\Psi$ is the scalar field. The coupling between the gravitation field and the scalar field is represented by the term $\xi R \Psi$, where $\xi$ is a numerical coupling factor. The normal mode solution can be expressed as
\begin{eqnarray}\label{S3}
\Psi_{\omega lm}=\frac{1}{h(r)}\chi_{\omega l}(r)Y_{lm}(\theta,\varphi)e^{-i\omega t},
\end{eqnarray}
where $Y_{lm}(\theta,\varphi)$ represents a scalar spherical harmonic on the unit two-sphere and $h(r)=\sqrt{r(r-2\epsilon)}$. We can easily obtain the radial equation
\begin{eqnarray}\label{S4}
\frac{d^{2}\chi_{\omega l}}{dr^{2}_{\ast}}+[\omega^{2}-V(r)]\chi_{\omega l}=0,
\end{eqnarray}
and
\begin{eqnarray}\label{S5}
V(r)=\frac{f(r)}{h(r)}\frac{d}{dr}\bigg[f(r)\frac{dh(r)}{dr}\bigg]+\frac{f(r)l(l+1)}{h^{2}(r)}+f(r)\bigg[\mu^{2}+\frac{2\xi \epsilon^{2}(r-2M)}{r^{2}(r-2\epsilon)^{3}}\bigg],
\end{eqnarray}
where $f(r)=(r-2M)/(r-2\epsilon)$ and the tortoise coordinate $r_{\ast}$ is defined by $dr_{\ast}=dr/f(r)$ \cite{SDF13}.

An analysis of Eq.(\ref{S4}) in the vicinity of the event horizon yields an incoming wave function that is analytic throughout the spacetime manifold
\begin{eqnarray}\label{S6}
\Psi_{\mathrm{in},\omega lm}=e^{-i\omega v}Y_{lm}(\theta,\varphi).
\end{eqnarray}
Furthermore, by performing calculations near the event horizon $r=r_{+}$ of the GHS black hole, one obtains the corresponding outgoing modes inside and outside the event horizon
\begin{eqnarray}\label{S7}
\Psi_{\mathrm{out},\omega lm}(r>r_{+})=e^{-i\omega u}Y_{lm}(\theta,\varphi),
\end{eqnarray}
\begin{eqnarray}\label{S8}
\Psi_{\mathrm{out},\omega lm}(r<r_{+})=e^{i\omega u}Y_{lm}(\theta,\varphi),
\end{eqnarray}
where $u=t-r_{\ast}$ and $v=t+r_{\ast}$. Eqs.(\ref{S7}) and (\ref{S8}) are separately analytic outside and inside the event horizon, and therefore constitute a complete orthogonal family.
By introducing the light-like Kruskal coordinates $U$ and $V$ via 
\begin{eqnarray}\label{S9}
u&=&-4(M-\epsilon)\ln[-U/(4M-4\epsilon)],\notag\\v&=&4(M-\epsilon)\ln[V/(4M-4\epsilon)],\mathrm{if}\;r>r_{+},\notag\\
u&=&-4(M-\epsilon)\ln[U/(4M-4\epsilon)],\notag\\v&=&4(M-\epsilon)\ln[V/(4M-4\epsilon)],\mathrm{if}\;r<r_{+},
\end{eqnarray}
the outgoing dilaton modes can be rewritten as
\begin{eqnarray}\label{S10}
\Phi_{\mathrm{out},\omega lm}(r>r_{+})=e^{4(M-\epsilon)i\omega\ln[U/(4M-4\epsilon)]}Y_{lm}(\theta,\varphi),
\end{eqnarray}
\begin{eqnarray}\label{S11}
\Phi_{\mathrm{out},\omega lm}(r<r_{+})=e^{-4(M-\epsilon)i\omega\ln[-U/(4M-4\epsilon)]}Y_{lm}(\theta,\varphi).
\end{eqnarray}
Following the Damour-Ruffini approach \cite{A3}, performing an analytic continuation of the modes in Eqs.(\ref{S10}) and (\ref{S11}) yields a complete set of positive‑frequency modes. Employing the identity $-1=e^{i\pi}$ and making Eq.(\ref{S10}) analytic in the lower half-plane of $U$, a complete basis of positive energy $U$ modes can be expressed as
\begin{eqnarray}\label{S12}
\Phi_{\mathrm{I},\omega lm}=e^{2\pi\omega(M-\epsilon)}\Phi_{\mathrm{out},\omega lm}(r>r_{+})
+e^{-2\pi\omega(M-\epsilon)}\Phi_{\mathrm{out},\omega lm}^{\ast}(r<r_{+}),
\end{eqnarray}
\begin{eqnarray}\label{S13}
\Phi_{\mathrm{II},\omega lm}=e^{-2\pi\omega(M-\epsilon)}\Phi_{\mathrm{out},\omega lm}^{\ast}(r>r_{+})
+e^{2\pi\omega(M-\epsilon)}\Phi_{\mathrm{out},\omega lm}(r<r_{+}).
\end{eqnarray}
Eqs.(\ref{S12}) and (\ref{S13}) are analytic for all real $U$ and $V$, and therefore provide a suitable basis for quantizing the field in Kruskal spacetime.

Performing second quantization of the dilaton field outside the black hole leads to a Bogoliubov transformation that connects the creation and annihilation operators in the dilaton and Kruskal spacetimes
\begin{eqnarray}\label{S14}
a_{K,\omega lm}^{\mathcal{B},\dagger}=\frac{1}{\sqrt{1-e^{-8\pi\omega(M-\epsilon)}}}b_{\mathrm{out},\omega lm}^{\mathcal{B},\dagger}
-\frac{1}{\sqrt{e^{8\pi\omega(M-\epsilon)}-1}}b_{\mathrm{in},\omega lm}^{\mathcal{B}},
\end{eqnarray}
\begin{eqnarray}\label{S15}
a_{K,\omega lm}^{\mathcal{B}}=\frac{1}{\sqrt{1-e^{-8\pi\omega(M-\epsilon)}}}b_{\mathrm{out},\omega lm}^{\mathcal{B}}-\frac{1}{\sqrt{e^{8\pi\omega(M-\epsilon)}-1}}b_{\mathrm{in},\omega lm}^{\mathcal{B},\dagger},
\end{eqnarray}
where $\mathcal{B}$ denotes the bosonic field. Here, $a_{K,\omega lm}^{\mathcal{B},\dagger}$ and $a_{K,\omega lm}^{\mathcal{B}}$ are the creation and annihilation operators acting on the Kruskal vacuum of the exterior region, while $b_{\mathrm{in},\omega lm}^{\mathcal{B},\dagger}$, $b_{\mathrm{in},\omega lm}^{\mathcal{B}}$ and $b_{\mathrm{out},\omega lm}^{\mathcal{B},\dagger}$, $b_{\mathrm{out},\omega lm}^{\mathcal{B}}$ are the corresponding operators acting on the dilaton vacuum of the interior and exterior regions, respectively. The Kruskal vacuum $|0\rangle_{K}^{\mathcal{B}}$ is defined by
$a_{K,\omega lm}^{\mathcal{B}}|0\rangle_{K}^{\mathcal{B}}=0. $
Upon normalizing the state vector, the Kruskal vacuum in dilaton spacetime is obtained as a maximally entangled two-mode squeezed state
\begin{eqnarray}\label{S17}
|0\rangle_{K}^{\mathcal{B}}=\sqrt{1-e^{-8\pi\omega(M-\epsilon)}}\sum^{\infty}_{n=0} e^{-4n\pi\omega(M-\epsilon)}|n\rangle_{\mathrm{out}}^{\mathcal{B}}|n\rangle_{\mathrm{in}}^{\mathcal{B}},
\end{eqnarray}
and the first excited state of the bosonic field reads
\begin{eqnarray}\label{S18}
|1\rangle_{K}^{\mathcal{B}}=a^{\mathcal{B},\dagger}_{K,\omega lm}|0\rangle_{K}^{\mathcal{B}}=[1-e^{-8\pi\omega(M-\epsilon)}]\sum^{\infty}_{n=0}\sqrt{n+1} e^{-4n\pi\omega(M-\epsilon)}|n+1\rangle_{\mathrm{out}}^{\mathcal{B}}|n\rangle_{\mathrm{in}}^{\mathcal{B}},
\end{eqnarray}
where $\{|n\rangle_{\mathrm{out}}\}$ and $\{|n\rangle_{\mathrm{in}}\}$ are the orthonormal bases for the exterior and interior regions of the event horizon, respectively \cite{SDF13,A4}. 
For an observer outside the dilaton black hole, the modes inside the event horizon lie in a causally disconnected region and are therefore inaccessible. After tracing over these interior degrees of freedom, the resulting Hawking radiation spectrum is obtained as
\begin{eqnarray}\label{S19}
N^{\mathcal{B}}_{\omega}=\frac{1}{e^{8\pi\omega(M-\epsilon)}-1}.
\end{eqnarray}
Eq.(\ref{S19}) shows that an observer in the exterior of the GHS dilaton black hole will detect a thermal Bose-Einstein distribution of particles when traversing the Kruskal vacuum. It should be noted that the quantization of a scalar field in a static, asymptotically flat black hole background is discussed in Appendix A.

\subsection{Fermionic field}
Following the same procedure as for the bosonic field, one obtains the Bogoliubov transformation between the Kruskal and dilaton operators for the fermionic field as
\begin{eqnarray}\label{S20}
a^{\mathcal{F},\dagger}_{K,\omega lm}=\frac{1}{\sqrt{e^{-8\pi\omega(M-\epsilon)}+1}}a^{\mathcal{F},\dagger}_{\mathrm{out},\omega lm}-\frac{1}{\sqrt{e^{8\pi\omega(M-\epsilon)}+1}}b_{\mathrm{in},\omega lm}^{\mathcal{F}},
\end{eqnarray}
\begin{eqnarray}\label{S21}
a^{\mathcal{F}}_{K,\omega lm}=\frac{1}{\sqrt{e^{-8\pi\omega(M-\epsilon)}+1}}a_{\mathrm{out},\omega lm}^{\mathcal{F}}-\frac{1}{\sqrt{e^{8\pi\omega(M-\epsilon)}+1}}b^{\mathcal{F},\dagger}_{\mathrm{in},\omega lm},
\end{eqnarray}
where the superscript $\mathcal{F}$ denotes the fermionic field \cite{B2,B3,B4}. Consequently, the Kruskal vacuum and the first excited state for the fermionic field in dilaton spacetime are given by
\begin{eqnarray}\label{S22}
|0\rangle_{K}^{\mathcal{F}}=\frac{1}{\sqrt{e^{-8\pi\omega(M-\epsilon)}+1}}|0\rangle_{\mathrm{out}}^{\mathcal{F}}|0\rangle_{\mathrm{in}}^{\mathcal{F}}+\frac{1}{\sqrt{e^{8\pi\omega(M-\epsilon)}+1}}|1\rangle_{\mathrm{out}}^{\mathcal{F}}|1\rangle_{\mathrm{in}}^{\mathcal{F}},
\end{eqnarray}
and
\begin{eqnarray}\label{S23}
|1\rangle_{K}^{\mathcal{F}}=|1\rangle_{\mathrm{out}}^{\mathcal{F}}|0\rangle_{\mathrm{in}}^{\mathcal{F}}.
\end{eqnarray}
Employing the same method used for the bosonic case, the resulting Hawking radiation spectrum for the fermionic field is
\begin{eqnarray}\label{S24}
N^{\mathcal{F}}_{\omega}=\frac{1}{e^{8\pi\omega(M-\epsilon)}+1},
\end{eqnarray}
which represents a thermal Fermi-Dirac distribution as measured by an observer outside the event horizon. 
Comparing this result with the bosonic spectrum in Eq.(\ref{S19}), it is clear that the distinct quantum statistics (Bose-Einstein versus Fermi-Dirac) lead to different gravitational effects in dilaton spacetime, which, in turn, influence quantum entanglement properties of the two fields in characteristically different ways.

\section{ One-tangle, two-Tangle and $\pi_4$-Tangle \label{GSCDGE}}
For a tetrapartite quantum state described by the density matrix $\rho_{ABCD}$, the entanglement among its subsystems satisfies the inequality
\begin{eqnarray}\label{S25}
N_{AB}^{2}+N_{AC}^{2}+N_{AD}^{2}\le N_{A(BCD)}^{2}, 
\end{eqnarray}
where $N_{AB}$, $N_{AC}$ and $N_{AD}$ are the two-tangles quantifying the bipartite entanglement between the respective pairs. Moreover, $N_{A(BCD)}$ is the one-tangle, which measures the entanglement between subsystem $A$ and the joint subsystem $BCD$. 
Formally, the one-tangle and two-tangle are defined via the negativity. For a subsystem $\alpha$ in a tetrapartite state $\rho_{\alpha\beta\gamma\delta}$, the one-tangle is defined as
\begin{eqnarray}\label{S26}
N_{\alpha (\beta \gamma \delta)}=\left\|\rho _{\alpha( \beta \gamma \delta )}^{T_{\alpha }}\right\|-1,
\end{eqnarray}
where $\rho _{\alpha( \beta \gamma \delta)}^{T_{\alpha }}$ denotes the partial transpose of the full tetrapartite density matrix with respect to subsystem $\alpha$ \cite{B5}. For a bipartite reduced state $\rho_{\alpha\beta}$, the two-tangle is defined as
\begin{eqnarray}\label{S27}
N_{\alpha \beta}=\left\|\rho _{\alpha \beta}^{T_{\alpha }}\right\|-1.
\end{eqnarray}
Here, $\rho _{\alpha \beta}^{T_{\alpha }}$ is the partial transpose of the reduced bipartite density matrix $\rho_{\alpha\beta}$, and $\left\|\rho\right\|=Tr\sqrt{\rho^{\dagger}\rho}$ is the trace norm. 
The quantity $\|\rho^{T_\alpha}\|-1$ equals twice the sum of the absolute values of the negative eigenvalues of the partially transposed matrix. Consequently, negativity can also be expressed as
\begin{eqnarray}\label{S28}
N(\rho )=2\sum_{\lambda _{i}< 0}\left |\lambda _{i}\right|,
\end{eqnarray}
where $\lambda_{i}$ is the eigenvalue of the partial transposed matrix.
As a widely adopted entanglement measure, negativity effectively quantifies the degree of quantum correlation between subsystems. The one-tangle is a suitable measure for detecting genuine multipartite entanglement. A quantum state possesses genuine multipartite entanglement if it cannot be separated under any bipartition, meaning its correlations involve all subsystems non-trivially. If the one-tangle of subsystem $\alpha$ vanishes ($N_{\alpha(\beta\gamma\delta)} = 0$), it indicates that $\alpha$ is separable from the rest of the system, and hence the state lacks genuine multipartite entanglement with respect to that partition.

The discrepancy between the two sides of the inequality Eq.(\ref{S25}) defines the residual entanglement for subsystem $A$ in a tetrapartite system
\begin{eqnarray}\label{S29}
\pi_{A}=N_{A(BCD)}^{2}-N_{AB}^{2}-N_{AC}^{2}-N_{AD}^{2}.
\end{eqnarray}
Similarly, the residual entanglement for the other subsystems are
\begin{eqnarray}\label{S30}
\pi _{B}=N_{B(ACD)}^{2}-N_{BA}^{2}-N_{BC}^{2}-N_{BD}^{2},
\end{eqnarray}
\begin{eqnarray}\label{S31}
\pi _{C}=N_{C(ABD)}^{2}-N_{CA}^{2}-N_{CB}^{2}-N_{CD}^{2},
\end{eqnarray}
\begin{eqnarray}\label{S311}
\pi _{D}=N_{D(ABC)}^{2}-N_{DA}^{2}-N_{DB}^{2}-N_{DC}^{2}.
\end{eqnarray}
In general, $\pi_{A}\ne\pi_{B}\ne\pi_{C}\ne\pi_{D}$, indicating that the residual entanglement is not invariant under permutation of the subsystems. For a tetrapartite system, an effective method to quantify multipartite global entanglement is to consider the average of the residual entanglement, known as the $\pi_{4}$-tangle. It is defined by
\begin{eqnarray}\label{S32}
\pi_{4}=\frac{1}{4}(\pi_{A}+\pi_{B}+\pi_{C}+\pi_{D}).
\end{eqnarray}
This measure captures the entanglement that is not accounted for by bipartite correlations alone, thus reflecting the intrinsic tetrapartite quantum correlation of the state. The quantity $\pi_{4}$ vanishes precisely when the entanglement of the system can be decomposed into purely bipartite contributions. Therefore, in this work, we adopt $\pi_{4}$-tangle to characterize the global entanglement properties of the tetrapartite system.

In this work, the entanglement is quantified by the negativity, which is particularly suitable for mixed states and has been widely adopted in relativistic quantum information. In the quantification of quantum multipartite entanglement, various approaches exist, each with certain limitations \cite{B9}. (1) Concurrence is widely used for bipartite systems, but it is not directly applicable to the multipartite systems considered in this work. (2) Entanglement of formation has a clear operational interpretation, yet its computation involves a complex optimization procedure that is difficult to carry out in practice. (3) Distillable entanglement characterizes the maximum amount of pure-state entanglement that can be extracted under LOCC, offering a well-defined resource interpretation; however, its evaluation is also highly nontrivial. (4) Squashed entanglement possesses favorable mathematical properties such as convexity, but its computation is extremely challenging. (5) Logarithmic negativity, as the logarithmic version of negativity, is convenient to compute, yet it does not satisfy LOCC monotonicity and therefore is not a full entanglement monotone. Furthermore, in the research on bosonic entanglement, negative values have been widely used as a measurement standard, while other entanglement metrics are less frequently employed \cite{SDF1,SDF2,SDF3,SDF4}. In summary, negativity strikes a favorable balance between computability and operational relevance, making it more suitable for the framework of this study.

\section{N-partite genuine entanglement of GHZ states for bosonic and fermionic fields in GHS dilaton spacetime\label{GSCDGE}}

The N-partite GHZ states for the bosonic and  fermionic  fields can be expressed as
\begin{eqnarray}\label{S33}
\left |GHZ\right \rangle _{N}^{\mathcal{B}/\mathcal{F}}=\frac{1}{\sqrt{2}}(\left |0\right \rangle^{\otimes N}+\left |1\right \rangle^{\otimes N}),
\end{eqnarray}
where $\mathcal{B}$ and $\mathcal{F}$  denote   bosonic and  fermionic  fields, respectively. 
We consider a scenario where  $m$ ($m<N$) particles are positioned near the event horizon of the GHS dilaton black hole, while the remaining  $N-m$ particles reside in the asymptotically flat region.

In this work, we adopt the standard framework of relativistic quantum information, in which the multipartite GHZ state is assumed to be initially prepared in an asymptotically flat region where all observers share a well-defined entangled state. Currently, the technology for preparing GHZ state in experiments is already quite mature \cite{B6,B7,B8}. After preparing the GHZ state, it is assumed that $N$ observers share the GHZ state, and that $m$ ($m<N$) observers are positioned near the event horizon of a black hole, while the remaining $N-m$ observers  reside in the asymptotically flat region. Due to the inequivalence between different field quantizations (e.g., Kruskal and dilaton modes), these observers perceive distinct vacuum structures, leading to the degradation and redistribution of quantum correlations. We emphasize that our analysis does not rely on the physical feasibility of preparing entanglement across the horizon at the time of observation, but rather focuses on how pre-existing entanglement is affected by gravitational effects. This setting follows the widely adopted paradigm in studies of quantum information in curved spacetime \cite{SDF1,SDF2,SDF3,SDF4,SDF5,SDF6,SDF7,SDF8,SDF9,SDF10,SDF11,SDF12,SDF13,SDF14,SDF15,SDF16,SDF17,SDF18,SDF19,SDF20,SDF21,SDF22}.

Using the dilaton modes of the bosonic field given in Eqs.(\ref{S17}) and (\ref{S18}), the state in Eq.(\ref{S33}) can be written as
\begin{eqnarray}\label{S34}
|GHZ\rangle_{N}^{\mathcal{B}}= &&\frac{1}{\sqrt{2}} \bigg\{\big[1-e^{-8\pi \omega(M-\epsilon)}\big]^{\frac{m}{2}}\big[\prod_{i=1}^{m} \sum_{n_{i}=0}^{\infty}e^{-4n_{i}\pi\omega(M-\epsilon)}\big]\big[\bigotimes_{i=1}^{m}|0\rangle^{\otimes q}\left|n_{i}\right\rangle_{\mathrm{out}}\left|n_{i}\right\rangle_{\mathrm{in}} \big]\notag\\
&&+\big[1-e^{-8\pi \omega(M-\epsilon)}\big
]^{m}\big[\prod_{i=1}^{m} \sum_{n_{i}=0}^{\infty}e^{-4n_{i}\pi\omega(M-\epsilon)}\sqrt{n_{i}+1}\big]\notag\\
&&\times\big[\bigotimes_{i=1}^{m} |1\rangle^{\otimes q}\left|n_{i}+1\right\rangle_{\mathrm{out}}\left|n_{i}\right\rangle_{\mathrm{in}}\big]\bigg\},
\end{eqnarray}
where $q=N-m$. Since the regions inside and outside the event horizon of the black hole are causally disconnected, we trace out the modes inside the event horizon to obtain the corresponding density operator
\begin{eqnarray}\label{S35}
\rho_{N}^{\mathcal{B}}(GHZ)&=&\frac{1}{2}\bigg\{\big[1-e^{-8\pi \omega(M-\epsilon)}\big]^{m}\big[\prod_{i=1}^{m}\sum_{n_{i}=0}^{\infty}e^{-8n_{i}\pi\omega(M-\epsilon)}\big]\big[\bigotimes_{i=1}^{m}|0\rangle^{\otimes q}\left|n_{i}\right\rangle\left\langle0\right|^{\otimes q}\left\langle n_{i}\right|\big]\notag\\
&+&\big[1-e^{-8\pi \omega(M-\epsilon)}\big]^{\frac{3m}{2}}\big[\prod_{i=1}^{m}\sum_{n_{i}=0}^{\infty}e^{-8n_{i}\pi\omega(M-\epsilon)}\sqrt{n_{i}+1}\big]\big[\bigotimes_{i=1}^{m}|0\rangle^{\otimes q}\left|n_{i}\right\rangle\notag\\
&\bigotimes&\left\langle1\right|^{\otimes q}\left\langle n_{i}+1\right|\big]+\big[1-e^{-8\pi \omega(M-\epsilon)}\big]^{\frac{3m}{2}}\big[\prod_{i=1}^{m}\sum_{n_{i}=0}^{\infty}e^{-8n_{i}\pi\omega(M-\epsilon)}\sqrt{n_{i}+1}\big]\notag\\
&\times&\big[\bigotimes_{i=1}^{m}|1\rangle^{\otimes q}\left|n_{i}+1\right\rangle\left\langle0\right|^{\otimes q}\left\langle n_{i}\right|\big]+\big[1-e^{-8\pi \omega(M-\epsilon)}\big]^{2m}\notag\\
&\times&\big[\prod_{i=1}^{m}\sum_{n_{i}=0}^{\infty}e^{-8n_{i}\pi\omega(M-\epsilon)}(n_{i}+1)\big]\big[\bigotimes_{i=1}^{m}|1\rangle^{\otimes q}\left|n_{i}+1\right\rangle\left\langle1\right|^{\otimes q}\left\langle n_{i}+1\right|\big]\bigg\}. 
\end{eqnarray}
For the sake of simplicity, we will omit the ``out" label from here on.

\begin{table}[htbp]
  \centering
  \includegraphics[width=0.6\textwidth]{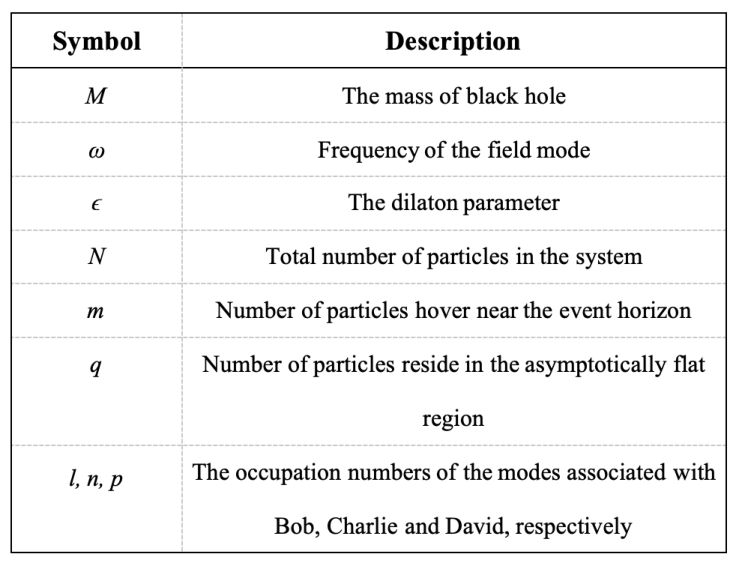}
  \caption{The symbol parameter table.}
  \label{Table 1}
\end{table}

We provide two summary tables of parameters (see Tables I and II) to clarify the physical meanings and interrelations of the key quantities used in the paper. The tables provide a systematic summary of all parameters, including their symbols and definitions, to clarify the notation used in the subsequent analysis. The parameters include the dilaton parameter $\epsilon$, the total number of particles in the system $N$, the number of particles located near the event horizon $m$, and the number of particles reside in the asymptotically flat region $q$, satisfying the relation $m+q=N$.

\begin{table}[htbp]
  \centering
  \includegraphics[width=0.6\textwidth]{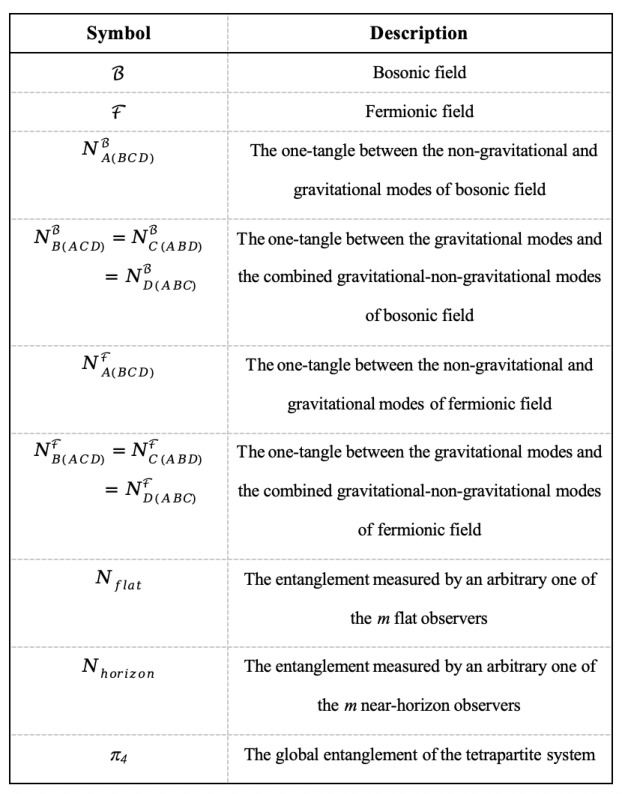}
  \caption{The entanglement symbol parameter table.}
  \label{Table 2}
\end{table}

The second table systematically summarizes the entanglement symbols used in this work. These include the one-tangle between the non-gravitational and gravitational modes for bosonic and fermionic fields, denoted as $N_{A(BCD)}^{\mathcal{B/F}}$, the one-tangle between the gravitational modes and the combined gravitational–non-gravitational modes, given by $N_{B(ACD)}^{\mathcal{B/F}}=N_{C(ABD)}^{\mathcal{B/F}}=N_{D(ABC)}^{\mathcal{B/F}}$, the entanglement measured by an arbitrary one of the $m$ flat observers, $N_{flat}$, the entanglement measured by an arbitrary one of the $m$ near-horizon observers, $N_{horizon}$ and the global entanglement of the tetrapartite system, $\pi_{4}$.

For concreteness, we now specialize to a tetrapartite GHZ state shared by four observers: Alice ($A$), Bob ($B$), Charlie ($C$), and David ($D$). We assume that the last three observers (Bob, Charlie, and David) are situated near the event horizon, while Alice remains in the asymptotically flat region. In this case ($N=4, m=3$), the density matrix in Eq.(\ref{S35}) reduces to
\begin{eqnarray}\label{S36}
\rho_{ABCD}^{\mathcal{B}}&=&\frac{1}{2}\sum_{l,n,p,=0}^{\infty}\bigg\{\big[1-e^{-8\pi \omega(M-\epsilon)}\big]^{3}e^{-8l\pi\omega(M-\epsilon)}e^{-8n\pi\omega(M-\epsilon)}e^{-8p\pi\omega(M-\epsilon)}|0lnp\rangle\left\langle0lnp\right|\notag\\
&+&\big[1-e^{-8\pi \omega(M-\epsilon)}\big]^{\frac{9}{2}}e^{-8l\pi\omega(M-\epsilon)}\sqrt{l+1}e^{-8n\pi\omega(M-\epsilon)}\sqrt{n+1}e^{-8p\pi\omega(M-\epsilon)}\sqrt{p+1}\notag\\
&\times &|0lnp\rangle\left\langle1(l+1)(n+1)(p+1)\right|+\big[1-e^{-8\pi \omega(M-\epsilon)}\big]^{\frac{9}{2}}e^{-8l\pi\omega(M-\epsilon)}\sqrt{l+1}\notag\\ &\times &e^{-8n\pi\omega(M-\epsilon)} \sqrt{n+1}e^{-8p\pi\omega(M-\epsilon)} \sqrt{p+1}|1(l+1)(n+1)(p+1)\rangle\left\langle0lnp\right| \notag\\ 
&+&\big[1-e^{-8\pi \omega(M-\epsilon)}\big]^{6} e^{-8l\pi\omega(M-\epsilon)}(l+1) e^{-8n\pi\omega(M-\epsilon)}(n+1)e^{-8p\pi\omega(M-\epsilon)}(p+1)\notag\\
&\times &|1(l+1)(n+1)(p+1)\rangle\left\langle1(l+1)(n+1)(p+1)\right|\bigg\},
\end{eqnarray}
where the indices $l,n,p$ label the occupation numbers of the modes associated with Bob, Charlie and David, respectively.
To compute the one-tangle for each observer, we construct the partial transpose of the density matrix with respect to each party. The corresponding transformation rules are
\begin{eqnarray}\label{S37}
&&|abcd\rangle \langle xyzw| \to |xbcd\rangle \langle ayzw|,\notag\\
&&|abcd\rangle \langle xyzw| \to |aycd\rangle \langle xbzw|,\notag\\
&&|abcd\rangle \langle xyzw| \to |abzd\rangle \langle xycw|,\notag\\
&&|abcd\rangle \langle xyzw| \to |abcw\rangle \langle xyyd|.
\end{eqnarray}

Since the product $(\rho_{ABCD}^{T_{A}})^{\dagger}\rho_{ABCD}^{T_{A}}$ is diagonal, the one‑tangle (negativity) for Alice, obtained from Eq.(\ref{S28}) is
\begin{eqnarray}\label{S38}	
N_{A(BCD)}^{\mathcal{B}}&=&\sum_{l,n,p=0}^{\infty }\sqrt{l+1}\big[1-e^{-8\pi\omega(M-\epsilon)}\big]^{\frac{3}{2}}e^{-8l\pi\omega(M-\epsilon)}\notag\\
&\times &\sqrt{n+1}\big[1-e^{-8\pi\omega(M-\epsilon)}\big]^{\frac{3}{2}}e^{-8n\pi\omega(M-\epsilon)}\notag\\
&\times &\sqrt{p+1}\big[1-e^{-8\pi\omega(M-\epsilon)}\big]^{\frac{3}{2}}e^{-8p\pi\omega(M-\epsilon)}.
\end{eqnarray}
When $\epsilon\to M$, we can obtain
\begin{eqnarray*}\label{SS38}	
&&\lim_{\epsilon \to M}\bigg\{\sum_{n=0}^{\infty }\sqrt{n+1}\big[1-e^{-8\pi\omega(M-\epsilon)}\big]^{\frac{3}{2}}e^{-8n\pi\omega(M-\epsilon)}\bigg\}=\frac{\sqrt{\pi}}{2}.
\end{eqnarray*}
Therefore, no numerical truncation is involved in evaluating the extremal limit.
For the numerical calculations away from the extremal limit, no closed-form expression is available for the infinite series in Eq.(\ref{S38}), and a truncation scheme is adopted. Throughout this work, the summation is truncated at $50000$. To verify numerical convergence, the cutoff is further increased to $9\times10^{10}$, and the numerical results remain unchanged to six decimal places. This demonstrates that the truncation error is negligible and does not affect any of the conclusions presented in this paper.

Owing to the permutation symmetry of the GHZ state among the three observers near the  event horizon, the one‑tangles for Bob, Charlie, and David are identical. They are given by
\begin{eqnarray}\label{S39}	
N_{B(ACD)}^{\mathcal{B}}=N_{C(ABD)}^{\mathcal{B}}=N_{D(ABC)}^{\mathcal{B}}=\sum_{l,n,p=0}^{\infty }\Big[\sqrt{(Y_{lnp}-Z_{lnp})^{2}+4F_{lnp}^{2}}-Y_{lnp}-Z_{lnp}\Big],
\end{eqnarray}
where the coefficients are
\begin{eqnarray}\label{S40}	
&&Y_{lnp}=\frac{\big[1-e^{-8\pi\omega(M-\epsilon)}\big]^{3}e^{-8(l+1)\pi\omega(M-\epsilon)}e^{-8(n+1)\pi\omega(M-\epsilon)}e^{-8(p+1)\pi\omega(M-\epsilon)}}{2},\notag\\
&&Z_{lnp}=\frac{\big[1-e^{-8\pi\omega(M-\epsilon)}\big]^{6}le^{-8(l-1)\pi\omega(M-\epsilon)}ne^{-8(n-1)\pi\omega(M-\epsilon)}pe^{-8(p-1)\pi\omega(M-\epsilon)}}{2},\notag\\
&&F_{lnp}=\frac{1}{2}\bigg\{\big[1-e^{-8\pi\omega(M-\epsilon)}\big]^{\frac{9}{2}}\sqrt{l+1} e^{-8l\pi\omega(M-\epsilon)}\sqrt{n+1}e^{-8n\pi\omega(M-\epsilon)}\notag\\
&&\times\sqrt{p+1}e^{-8p\pi\omega(M-\epsilon)}\bigg\}.
\end{eqnarray}
It should be noted that all calculations regarding fermionic entanglement can be found in Appendix B.

\begin{figure}
\begin{minipage}[t]{0.5\linewidth}
\centering
\includegraphics[width=3.2in,height=6.5cm]{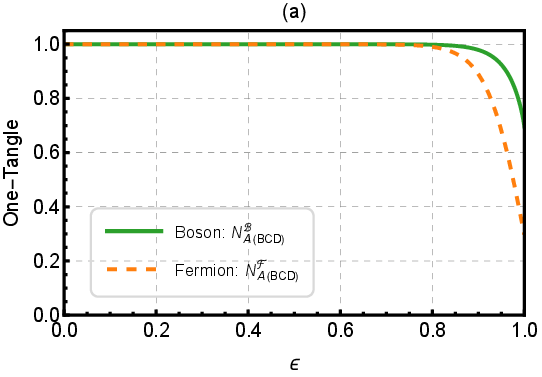}
\label{fig1aa}
\end{minipage}%
\begin{minipage}[t]{0.5\linewidth}
\centering
\includegraphics[width=3.2in,height=6.5cm]{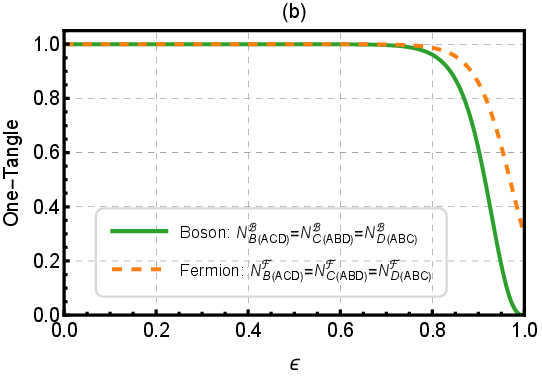}
\label{fig1bb}
\end{minipage}%
\caption{The one-tangle of the GHZ state for bosonic and fermionic fields as a function of the dilaton $\epsilon$ for the fixed values $M=\omega=1$.}
\label{Fig1}
\end{figure}

As shown in Fig.\ref{Fig1}, we present the one-tangle of the tetrapartite GHZ state for both bosonic and fermionic fields as a function of the dilaton  $\epsilon$, with $M=\omega=1$ fixed. 
For both types of fields, the one-tangle decreases monotonically with increasing dilaton $\epsilon$, indicating that the Hawking effect induced by the dilaton black hole progressively degrades quantum entanglement. This monotonic suppression is observed in both bipartitions considered: the entanglement between the non-gravitational and gravitational modes [Fig.\ref{Fig1}(a)], and that between the gravitational modes and the combined gravitational-non-gravitational modes [Fig.\ref{Fig1}(b)]. 
Despite this common qualitative behavior, a pronounced distinction emerges between bosonic and fermionic fields in curved spacetime. In Fig.\ref{Fig1}(a), which corresponds to the entanglement between the non-gravitational and gravitational modes, the bosonic field exhibits stronger entanglement than its fermionic counterpart in  dilaton spacetime. This ordering is opposite to that typically encountered in conventional bipartite systems, where fermionic entanglement is generally more robust than bosonic entanglement \cite{SDF1,SDF2,SDF3,SDF4,SDF5,SDF6,SDF7,SDF8}.  Such a reversal challenges the widely held intuition regarding the superiority of fermionic entanglement and highlights the nontrivial role played by strong gravitational effects in reshaping entanglement structures.
By contrast, for the entanglement between the gravitational modes and the combined 
gravitational-non-gravitational modes shown in Fig.\ref{Fig1}(b), the bosonic field displays weaker entanglement than the fermionic field once the dilaton black hole is present.  These contrasting behaviors demonstrate that the relative robustness of bosonic and fermionic entanglement depends sensitively on both the spacetime background and the specific mode partition considered.
Overall, our results provide a fresh perspective on quantum entanglement in strong gravitational fields and offer valuable insights into the selection, optimization, and protection of quantum resources in relativistic and curved-spacetime quantum information tasks.
 
The distinct behavior between bosonic and fermionic entanglement can be understood from their fundamentally different statistical properties and Hilbert space structures. For fermionic fields, the Pauli exclusion principle restricts each mode to a finite-dimensional occupation (at most one particle per state), which effectively limits the degree of mode mixing induced by Hawking radiation. As a result, although entanglement is degraded by the gravitational field, a finite amount of quantum correlation can remain even in the strong-gravity regime. In contrast, bosonic fields allow unbounded occupation numbers, leading to an infinite-dimensional Hilbert space. This feature enables stronger mode mixing between accessible and inaccessible regions as the dilaton parameter increases, which in turn redistributes quantum correlations more drastically. In the weak-gravity regime, this enhanced mixing tends to accelerate the degradation of entanglement for bosons. However, in the strong-gravity regime, the same mechanism can lead to a relative enhancement of residual correlations compared to the fermionic case, resulting in the observed crossover where bosonic entanglement exceeds fermionic entanglement. Therefore, the competition between bounded fermionic statistics and unbounded bosonic excitations provides a natural explanation for the different responses of entanglement to the Hawking effect.

To compute the global entanglement $\pi_{4}$  according to Eq.(\ref{S32}), we first need to evaluate the residual entanglement  \(\pi_A, \pi_B, \pi_C, \pi_D\). When analyzing the GHZ state, we find that the reduced-state entanglement between any two particles is zero; that is, for every particle pair \(\alpha\beta\) (e.g., \(AB, AC, AD, BC, BD,\) or \(CD\)), the bipartite entanglement measure satisfies \(N_{\alpha\beta}=0\). Due to this property, all residual entanglements  \(\pi_A, \pi_B, \pi_C, \pi_D\)  in  Eqs.(\ref{S29})-(\ref{S311}) consequently reduce to a much simpler form
\begin{eqnarray}\label{SS41}
&&\pi_{A}=N_{A(BCD)}^{2}, \; \pi_{B}=N_{B(ACD)}^{2},\notag\\ 
&&\pi_{C}=N_{C(ABD)}^{2}, \; \pi_{D}=N_{D(ABC)}^{2}.
\end{eqnarray}
Therefore,  the global entanglement $\pi_{4}$ in  Eq.(\ref{S32})  can be expressed as
\begin{eqnarray}\label{S41}
\pi_{4}=\frac{1}{4}\Big[N_{A(BCD)}^{2}+N_{B(ACD)}^{2}+N_{C(ABD)}^{2}+N_{D(ABC)}^{2}\Big].
\end{eqnarray}

\begin{figure}
\begin{minipage}[t]{0.5\linewidth}
\centering
\includegraphics[width=3.5in,height=6.5cm]{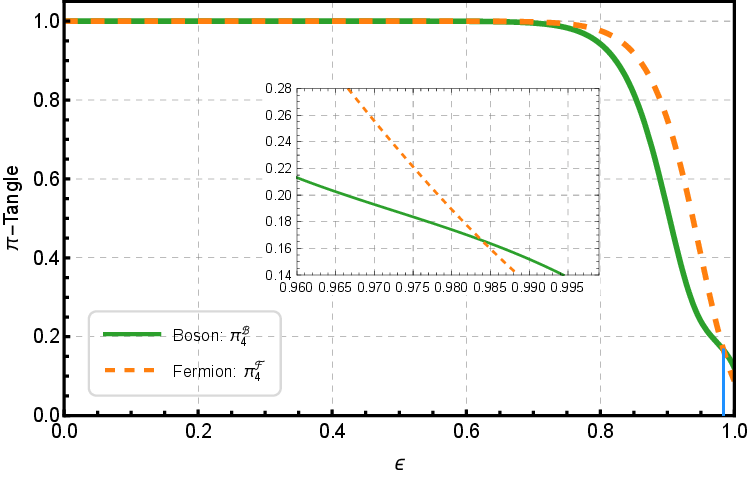}
\label{fig5}
\end{minipage}%
\caption{Tetrapartite $\pi$-tangle of bosonic and fermionic fields versus the dilaton $\epsilon$, with fixed $M=\omega=1$.}
\label{Fig5}
\end{figure}

Fig.\ref{Fig5} displays the tetrapartite $\pi$-tangle of bosonic and fermionic fields as a function of the dilaton  $\epsilon$, with $M=\omega=1$ fixed. 
For both field types, the $\pi$-tangle decreases monotonically as the dilaton $\epsilon$ increases, reflecting the progressive degradation of multipartite entanglement induced by the Hawking effect in a dilaton black hole background. Beyond this overall suppression, a detailed comparison reveals a pronounced crossover behavior between the bosonic and fermionic cases. 
In the regime of small $\epsilon$, corresponding to a weak gravitational background, the fermionic field exhibits a larger $\pi$-tangle than the bosonic field.  As $\epsilon$ increases and the gravitational field becomes strong, this ordering is reversed: the bosonic
$\pi$-tangle surpasses that of the fermionic field and eventually becomes dominant. This crossover suggests a transition in the underlying physical mechanism governing multipartite entanglement. 
While particle statistics dominate the entanglement hierarchy in weak gravitational fields, strong spacetime curvature and enhanced Hawking radiation progressively reshape the entanglement structure, favoring the sustainability of bosonic multipartite correlations over their fermionic counterparts. 
Consequently, the bosonic system demonstrates greater robustness against gravitationally induced decoherence in the strong-gravity regime. In the present work, the term ``strong gravity" refers to the regime of enhanced spacetime curvature and Hawking-induced mode mixing near the black-hole horizon. For the GHS spacetime, this regime corresponds to the near-extremal region, where the dilaton parameter approaches its extremal value $M$ (for example, $\epsilon=0.99M$). Therefore, the stronger robustness of bosonic multipartite entanglement reported here should be understood in this specific sense. Moreover, in the extremal limit $\epsilon \to M$, the residual multipartite entanglement approaches a finite value that is independent of the mode frequency $\omega$, indicating that the influence of the field frequency becomes negligible compared with the dominant gravitational effects in this regime.
From the perspective of relativistic quantum information, these results provide important guidance for the selection of quantum resources in curved spacetime. 
In particular, they indicate that fermionic multipartite states may be more advantageous for quantum information tasks in weakly relativistic settings, whereas bosonic multipartite resources could offer superior performance and resilience in strong gravitational environments.

The result of Eq.(\ref{S38}) can be extended to a more general N-partite systems. Specifically, for a multipartite GHZ state with $m$ observers located near the event horizon of the dilaton black hole, the entanglement accessible to the remaining observers in the asymptotically flat region is given by
\begin{eqnarray}\label{S42}	
N_{\mathrm{flat}}^{\mathcal{B}}=\big[1-e^{-8\pi\omega(M-\epsilon)}\big]^{\frac{3m}{2}}\Bigg[\prod_{i=1}^{m}\sum_{n_{i}=0}^{\infty }\sqrt{n+1_{i}}e^{-8n_{i}\pi\omega(M-\epsilon)}\Bigg].
\end{eqnarray}
Similarly, based on Eq.(\ref{S39}), the entanglement measured by the $m$ observers near the horizon can be written as
\begin{eqnarray}\label{S42}	
N_{\text {horizon}}^{\mathcal{B}}=\sum_{n_{1}, \ldots, n_{m}=0}^{\infty}\Bigg[\sqrt{\left(Y_{n_{1}, \ldots, n_{m}}-Z_{n_{1}, \ldots, n_{m}}\right)^{2}+4 F_{n_{1}, \ldots, n_{m}}^{2}}-Y_{n_{1}, \ldots, n_{m}}-Z_{n_{1}, \ldots, n_{m}}\Bigg],
\end{eqnarray}
where $Y_{n_{1},\dots,n_{m}}$, $F_{n_{1},\dots,n_{m}}$, and $Z_{n_{1},\dots,n_{m}}$ are given by
\begin{eqnarray}\label{S43}	
&&Y_{n_{1},\dots,n_{m}}=\frac{\big[1-e^{-8\pi\omega(M-\epsilon)}\big]^{m}\big[\prod_{i=1}^{m}e^{-8(n_{i}+1)\pi\omega(M-\epsilon)}\big]}{2},\notag\\
&&Z_{n_{1},\dots,n_{m}}=\frac{\big[1-e^{-8\pi\omega(M-\epsilon)}\big]^{2m}\big[\prod_{i=1}^{m}n_{i}e^{-8(n_{i}-1)\pi\omega(M-\epsilon)}\big]}{2},\notag\\
&&F_{n_{1},\dots,n_{m}}=\frac{\big[1-e^{-8\pi\omega(M-\epsilon)}\big]^{\frac{3m}{2}}\big[\prod_{i=1}^{m}\sqrt{n+1_{i}}e^{-8n_{i}\pi\omega(M-\epsilon)}\big]}{2}.
\end{eqnarray}

\begin{figure}
\begin{minipage}[t]{0.5\linewidth}
\centering
\includegraphics[width=3.2in,height=6.5cm]{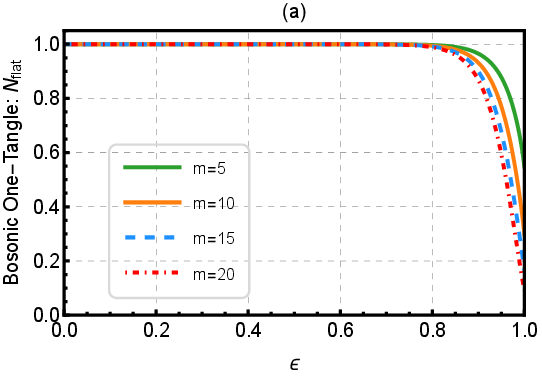}
\label{fig1aa}
\end{minipage}%
\begin{minipage}[t]{0.5\linewidth}
\centering
\includegraphics[width=3.2in,height=6.5cm]{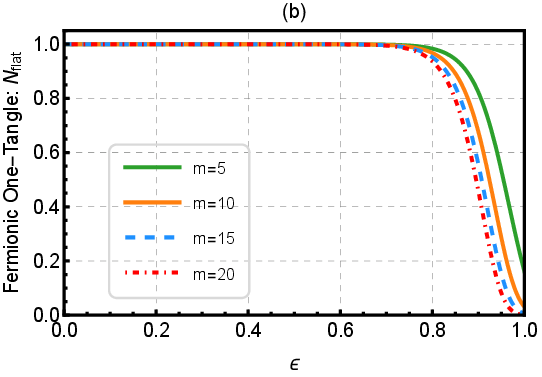}
\label{fig1bb}
\end{minipage}%
\caption{N‑partite one‑tangle of the GHZ state between the non‑gravitational and gravitational modes for bosonic and fermionic fields as a function of the dilaton $\epsilon$, with $M=\omega=1$. }
\label{Fig3}
\end{figure}

In Fig.\ref{Fig3}, we plot the N-partite one-tangle between the non-gravitational and gravitational modes of a GHZ state for bosonic and fermionic fields as a function of the dilaton  $\epsilon$, with $M=\omega=1$ fixed. 
For both fields, the entanglement remains approximately stable at small $\epsilon$, but then decreases rapidly as $\epsilon$ approaches $M$. 
Notably, the bosonic field's entanglement never vanishes completely and remains consistently higher than that of the fermionic field. A key observation is that the decay of quantum entanglement accelerates significantly as the number of included gravitational modes increases, driving the entanglement more rapidly toward zero.
This behavior can be understood as a consequence of the multipartite GHZ state's sensitivity to local noise and the structure of Hawking-induced decoherence: each gravitational mode interacts with the inaccessible modes inside the event horizon, effectively acting as an additional noisy environment. 
Tracing out the inaccessible modes distributes the original entanglement across more subsystems, thereby reducing the amount of entanglement observable in the accessible modes. 
As a result, the larger the number of participating gravitational modes, the greater the cumulative decoherence effect, leading to a faster suppression of the one-tangle. 
Moreover, due to the finite-dimensional Hilbert space of fermionic modes, this degradation is particularly pronounced for fermionic fields, whereas the infinite-dimensional nature of bosonic modes allows a residual entanglement to persist even under strong gravitational effects. These findings indicate that in the dilaton black hole background, the  entanglement depends not only on the type of field (bosonic or fermionic) but also sensitively on the number of gravitational modes included, providing important insights into the robustness of quantum resources in curved spacetime and guiding their selection for relativistic quantum information tasks.

\begin{figure}
\begin{minipage}[t]{0.5\linewidth}
\centering
\includegraphics[width=3.2in,height=6.5cm]{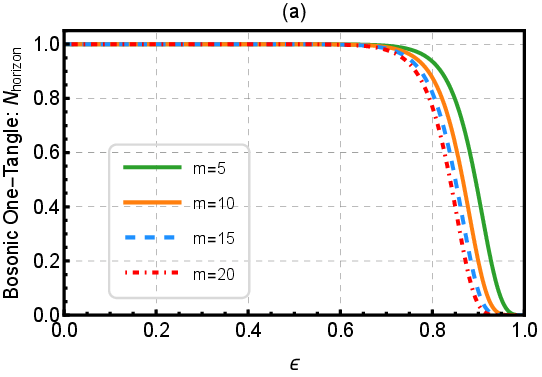}
\label{fig1aa}
\end{minipage}%
\begin{minipage}[t]{0.5\linewidth}
\centering
\includegraphics[width=3.2in,height=6.5cm]{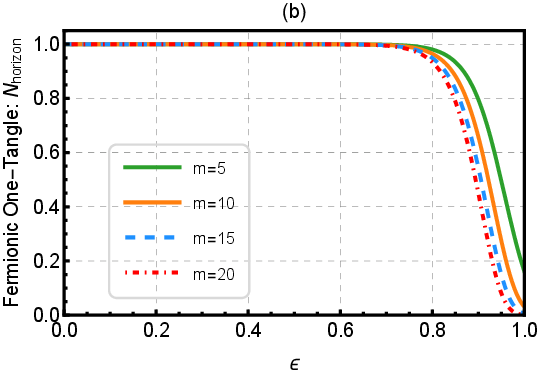}
\label{fig1bb}
\end{minipage}%
\caption{N‑partite one‑tangle of the GHZ state between the gravitational modes and the combined gravitational and non-gravitational modes for bosonic and fermionic fields as a function of the dilaton $\epsilon$, with $M=\omega=1$. }
\label{Fig4}
\end{figure}

In Fig.\ref{Fig4}, we present the $N$-partite one-tangle between the gravitational modes and the combined gravitational and non-gravitational modes  for bosonic and fermionic fields as a function of the dilaton $\epsilon$, with $M=\omega=1$ fixed. 
The overall behavior closely resembles that observed in Fig.\ref{Fig3}: for both fields, the entanglement remains nearly unchanged in the weak-gravity regime of small $\epsilon$, and then decays rapidly as $\epsilon$ approaches $M$. A notable qualitative difference, however, arises in the relative robustness of the two fields. 
In contrast to Fig.\ref{Fig3}, where the bosonic field retains stronger entanglement, the fermionic entanglement in the present bipartition remains globally higher than that of the bosonic field in dilaton spacetime. 
This reversal highlights the strong partition dependence of quantum entanglement in curved spacetime and underscores the nontrivial role played by particle statistics in the presence of gravitationally induced decoherence. These findings emphasize that in relativistic quantum information processing, the robustness of quantum resources is not determined solely by the field type, but also by the specific mode partition under consideration.

\begin{figure}
\begin{minipage}[t]{0.5\linewidth}
\centering
\includegraphics[width=3.0in,height=5.2cm]{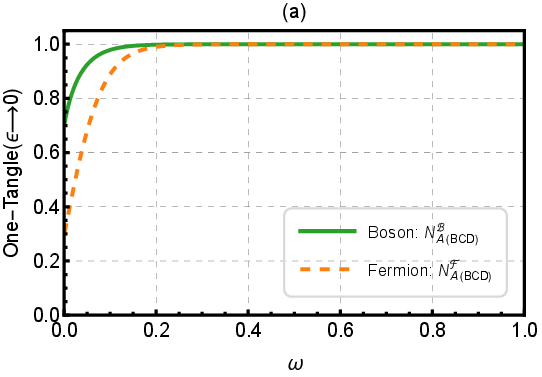}
\label{fig6a}
\end{minipage}%
\begin{minipage}[t]{0.5\linewidth}
\centering
\includegraphics[width=3.0in,height=5.2cm]{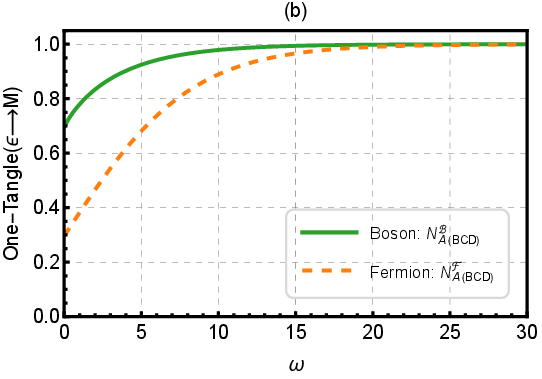}
\label{fig6b}
\end{minipage}%

\begin{minipage}[t]{0.5\linewidth}
\centering
\includegraphics[width=3.0in,height=5.2cm]{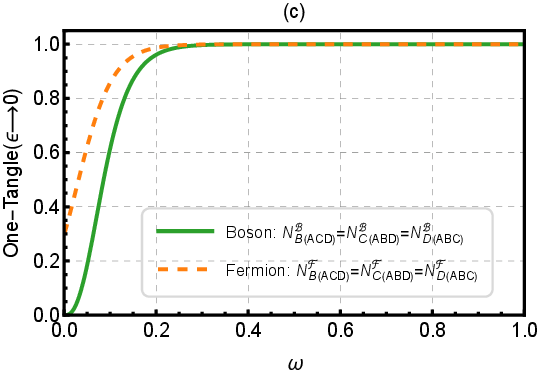}
\label{fig6c}
\end{minipage}%
\begin{minipage}[t]{0.5\linewidth}
\centering
\includegraphics[width=3.0in,height=5.2cm]{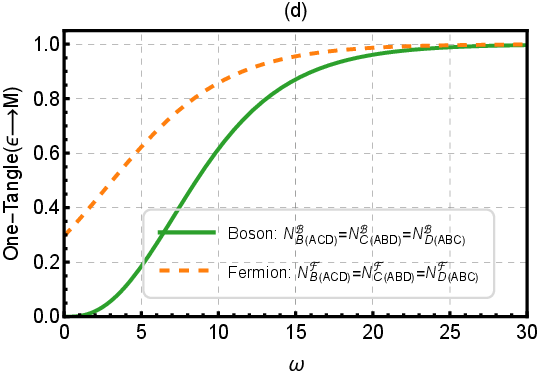}
\label{fig6d}
\end{minipage}%

\begin{minipage}[t]{0.5\linewidth}
\centering
\includegraphics[width=3.0in,height=5.2cm]{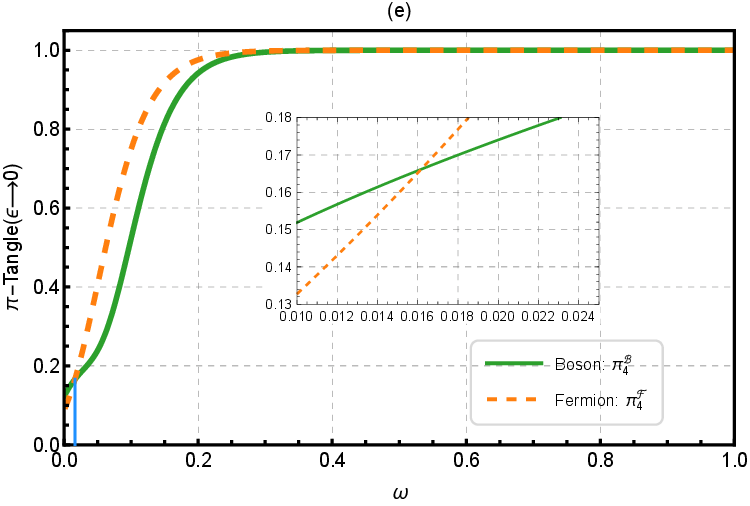}
\label{fig6e}
\end{minipage}%
\begin{minipage}[t]{0.5\linewidth}
\centering
\includegraphics[width=3.0in,height=5.2cm]{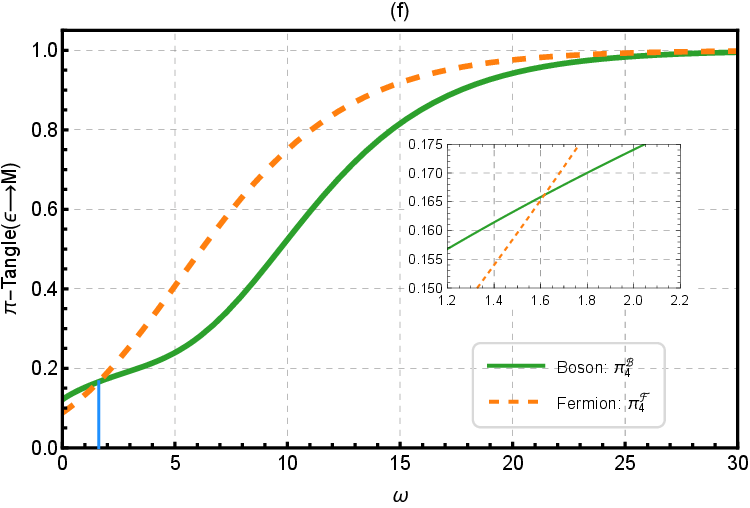}
\label{fig6f}
\end{minipage}%
\caption{The one-tangle and $\pi$-tangle for the bosonic and fermionic fields are plotted as functions of the frequency $\omega$ in two gravity limits, with $M=1$ fixed. It is noted that the three figures in the left column, with $\epsilon=0.001$ approximately correspond to the weak-gravity limit $(\epsilon \to 0)$, while the three figures in the right column, with $\epsilon=0.99$ approximately correspond to the strong-gravity limit $(\epsilon \to M)$.}
\label{Fig6}
\end{figure}
In Fig.\ref{Fig6}, we plot one-tangle and $\pi$-tangle between the non-gravitational and gravitational modes with between the gravitational modes and the combined gravitational and non-gravitational modes for bosonic and fermionic fields as a function of the frequency $\omega$, in the strong-gravity region $(\epsilon =0.99)$ and the weak-gravity region $(\epsilon=0.001)$. In both the weak-gravity and strong-gravity regions, the one-tangle and the $\pi$-tangle increase with $\omega$. As can be seen from the figures, for the one-tangle between non-gravitational and gravitational modes, the bosonic field exhibits stronger entanglement than the fermionic field; while for the one-tangle between gravitational modes and the combined system, the fermionic field exhibits stronger entanglement than the bosonic field. It is worth noting that the crossing behavior of the $\pi$-tangle (i.e., the reversal of the relative strength between the bosonic and fermionic fields) persists over a wide range of frequencies, and as $\omega$  increases, the crossing point shifts slightly toward larger $\epsilon$. This confirms that the main conclusion regarding the crossing behavior of global multipartite entanglement is robust against variations in the field parameter $\omega$.

\section{ Conclusions  \label{GSCDGE}}
In this paper, we have investigated the genuine N-partite entanglement of bosonic and fermionic GHZ states in the presence of a GHS dilaton black hole, using negativity as the entanglement measure. Our study began by considering a scenario in which $N$ observers share N-partite GHZ states in an asymptotically flat region, followed by the introduction of a gravitational environment. In this environment, $N-m$ observers remain inertial in flat spacetime, while $m$ observers are positioned near the event horizon of the dilaton black hole.
Our findings reveal several important physical insights: \textbf{(i) reversal of bosonic and fermionic entanglement in non-gravitational-gravitational partitions:}  the quantum entanglement between non-gravitational and gravitational modes is stronger in the bosonic field than in the fermionic field within dilaton spacetime. These results challenge the traditional notion that fermionic entanglement consistently outperforms bosonic entanglement for bipartite systems  in the relativistic framework \cite{SDF1,SDF2,SDF3,SDF4,SDF5,SDF6,SDF7,SDF8}; 
\textbf{ (ii) partition-dependent dominance of fermionic entanglement near the event horizon:} however, when examining the entanglement between the gravitational modes and the combined gravitational and non-gravitational modes, the fermionic field shows greater entanglement than the bosonic field in the presence of the dilaton black hole; \textbf{ (iii) nontrivial crossover of global multipartite entanglement with gravitational strength:}
the relative magnitude of the global multipartite entanglement in bosonic and fermionic fields is not fixed but evolves nontrivially with the dilaton. 
As the gravitational strength increases, the bosonic multipartite entanglement is initially weaker than its fermionic counterpart; however, beyond a certain threshold, this ordering is reversed and the bosonic entanglement becomes dominant. 
This nonmonotonic interchange indicates that the gravitational field strength plays a decisive role in shaping the hierarchy of multipartite entanglement between different quantum fields.  We emphasize that the present results are obtained within the framework of negativity, and while the observed crossover behavior is expected to be qualitatively robust, a systematic analysis using alternative entanglement measures remains an important direction for future investigation.

From a relativistic quantum information perspective, these findings emphasize that the suitability of quantum resources depends sensitively on the gravitational environment. 
Fermionic multipartite states are better suited for information-processing tasks in weak gravitational fields, whereas bosonic multipartite resources offer enhanced robustness and operational advantages in strongly curved spacetimes. 
Overall, our results expose the intricate, field-dependent response of quantum entanglement to gravity, highlighting how spacetime conditions can fundamentally reshape the structure and resilience of quantum correlations and opening new avenues for optimizing and protecting entanglement in relativistic and gravitational settings.

\begin{acknowledgments}
This work is supported by the National Natural Science Foundation of China (Grant nos. 12175095 and 12575056 
), the Natural
Science Foundation of the Liaoning Scientific Committee No. 2026-MS-281, and LiaoNing Revitalization Talents Program (XLYC2007047 and XLYC2503099).
\end{acknowledgments}

\appendix
\onecolumngrid
%


\appendix
\section{ Vacuum structure and Hawking radiation of scalar field}
It is well known that the spherically symmetric line element of a static, asymptotically flat black hole such as the Schwarzschild black hole, the GHS dilaton black hole \cite{A3}, the Reissner-Nordstr\"{o}m black hole \cite{B10},  the Casadio-Fabbri-Mazzacurati brane black hole \cite{B11}, and so on can be written in the form
\begin{eqnarray}\label{SS1}
ds^{2}=f(r)dt^{2}-\frac{1}{h(r)}dr^{2}-R^{2}(r)(d\theta^{2}+\sin{\theta}^{2}d\varphi^{2}),
\end{eqnarray}
where the functions $f(r)$ and $h(r)$ both vanish at the event horizon $r=r_{+}$. The surface gravity at the horizon is then given by $\kappa=\sqrt{f'(r_{+})h'(r_{+})}/2$. Introducing the tortoise coordinate $r_{\ast}$ via $dr_{\ast}=dr/\sqrt{f(r)h(r)}$, the metric Eq.(\ref{SS1}) can be expressed as
\begin{eqnarray}\label{SS2}
ds^{2}=f(r)(dt^{2}-dr_{\ast}^{2})-R^{2}(r)(d\theta^{2}+\sin{\theta}^{2}d\varphi^{2}).
\end{eqnarray}

The massless scalar field $\psi$ satisfies the Klein-Gordon equation
\begin{eqnarray}\label{SS3}
\frac{1}{\sqrt{-g}}\frac{\partial}{\partial x^{\mu}}(\sqrt{-g}g^{\mu\nu}\frac{\partial\psi }{\partial x^{\nu}})=0.
\end{eqnarray}
After expressing the normal mode solution as \cite{B12,B13}
\begin{eqnarray}\label{SS4}
\psi_{\omega lm}=\frac{1}{R(r)}\chi_{\omega l}(r)Y_{lm}(\theta,\varphi)e^{-i\omega t},
\end{eqnarray}
we can easily get the radial equation
\begin{eqnarray}\label{SS5}
\frac{d^{2}\chi_{\omega l}}{dr^{2}_{\ast}}+[\omega^{2}-V(r)]\chi_{\omega l}=0,
\end{eqnarray}
with
\begin{eqnarray}\label{SS6}
V(r)=\frac{\sqrt{f(r)h(r)}}{R(r)}\frac{d}{dr}[\sqrt{f(r)h(r)}\frac{dR(r)}{dr}]+\frac{l(l+1)f(r)}{R^{2}(r)},
\end{eqnarray}
where $Y_{lm}(\theta,\varphi)$ denotes a scalar spherical harmonic on the unit twosphere.

By performing a series of derivations near the event horizon, one can introduce the generalized lightlike Kruskal coordinates \cite{B13,B14,B15,B16}
\begin{eqnarray}\label{SS12}
&&U=-\frac{1}{\kappa}e^{-\kappa u},V=\frac{1}{\kappa}e^{-\kappa v}, \mathrm{if} r>r_{+};\notag\\
&&U=\frac{1}{\kappa}e^{-\kappa u},V=\frac{1}{\kappa}e^{-\kappa v}, \mathrm{if} r<r_{+},
\end{eqnarray}
and noticing that near the event horizon
\begin{eqnarray}\label{SS13}
r_{\ast} \simeq \frac{1}{\sqrt{f'(r_{+})h'(r_{+})}}\ln{(r-r_{+}}).
\end{eqnarray}

We adopt the gravitational mode and the Kruskal mode to describe the field. Therefore, the Kruskal vacuum $|0\rangle_{K}^{\mathcal{B}}$ is related to the vacuum of the black hole $|0\rangle_{\mathrm{out}}^{\mathcal{B}}\otimes|0\rangle_{\mathrm{in}}^{\mathcal{B}}$ by
\begin{eqnarray}\label{SS14}
|0\rangle^{\mathcal{B}}_{K}=\Upsilon(b_{\mathrm{out},\omega lm}^{\mathcal{B}},b_{\mathrm{out},\omega lm}^{{\mathcal{B}},\dagger},b_{\mathrm{in},\omega lm}^{\mathcal{B}},b_{\mathrm{in},\omega lm}^{{\mathcal{B}},\dagger})|0\rangle_{\mathrm{out}}^{\mathcal{B}}\otimes|0\rangle_{\mathrm{in}}^{\mathcal{B}}.
\end{eqnarray}
From $[b_{\mathrm{out},\omega lm}^{\mathcal{B}},b_{\mathrm{out},\omega lm}^{{\mathcal{B}},\dagger}]=[b_{\mathrm{in},\omega lm}^{\mathcal{B}},b_{\mathrm{in},\omega lm}^{{\mathcal{B}},\dagger}]=1$, we get \cite{B12,B18}
\begin{eqnarray}\label{SS15}
\Upsilon\propto \exp{(b_{\mathrm{out},\omega lm}^{{\mathcal{B}},\dagger}b_{\mathrm{in},\omega lm}^{{\mathcal{B}},\dagger}e^{-\pi\omega/\kappa})}.
\end{eqnarray}
After properly normalizing the state vector, we obtain the Kruskal vacuum which is a maximally entangled two-mode squeezed state \cite{B17,B18}
\begin{eqnarray}\label{SS16}
|0\rangle^{\mathcal{B}}_{K}=\sqrt{1-e^{-2\pi\omega/\kappa}}\sum_{n=0}^{\infty}e^{-n\pi\omega/\kappa}|n\rangle_{\mathrm{out}}^{\mathcal{B}}\otimes|n\rangle_{\mathrm{in}}^{\mathcal{B}},
\end{eqnarray}
and the first excited state
\begin{eqnarray}\label{SS17}
|1\rangle^{\mathcal{B}}_{K}&&=a^{{\mathcal{B}},\dagger}_{\mathrm{out},\omega lm}|0^{\mathcal{B}}_{K}\rangle\notag\\
&&=(1-e^{-2\pi\omega/\kappa})\sum_{n=0}^{\infty}\sqrt{n+1}e^{-n\pi\omega/\kappa}|n+1\rangle_{\mathrm{out}}^{\mathcal{B}}\otimes|n\rangle_{\mathrm{in}}^{\mathcal{B}},
\end{eqnarray}
where $\{|n\rangle^{\mathcal{B}}_{\mathrm{out}}\}$ and $\{|n\rangle^{\mathcal{B}}_{\mathrm{in}}\}$ are the orthonormal bases for the outside region and inside region of the event horizon, respectively. Here, the Hawking temperature can be defined as \cite{B19,B20}
\begin{eqnarray}\label{SS19}
T=\frac{\kappa}{2\pi}=\frac{\sqrt{f'(r_{+})h'(r_{+})}}{4\pi}.
\end{eqnarray}
The derivation for the fermionic field is analogous to that for the bosonic field and is omitted here to avoid redundancy.

\begin{figure}
\begin{minipage}[t]{0.5\linewidth}
\centering
\includegraphics[width=3.2in,height=6.5cm]{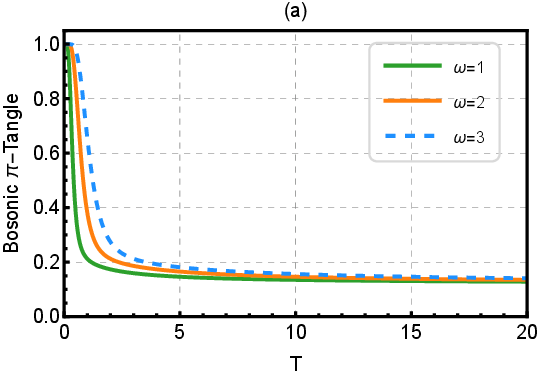}
\label{fig1aa}
\end{minipage}%
\begin{minipage}[t]{0.5\linewidth}
\centering
\includegraphics[width=3.2in,height=6.5cm]{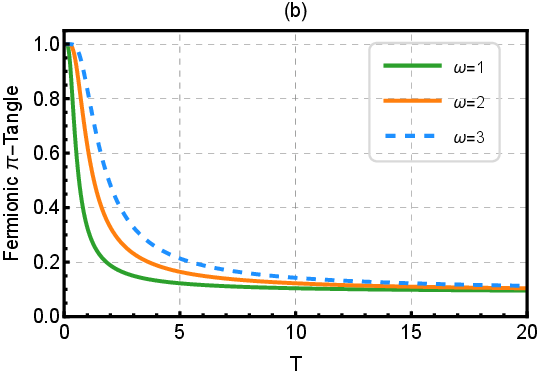}
\label{fig1bb}
\end{minipage}%
\caption{Tetrapartite $\pi$-tangle of the GHZ state for bosonic and fermionic fields as a function of the Hawking temperature $T$ for different $\omega$ in Schwarzschild spacetime.}
\label{Fig8}
\end{figure}
To determine whether the observed crossover behavior (bosonic entanglement surpassing fermionic entanglement in the strong‑field regime) is an artifact of the special extremal geometry of the GHS black hole, we perform an independent analysis for the Schwarzschild black hole. Considering the Hawking temperature $T=\frac{1}{8\pi M}$ in the Schwarzschild background, we plot the tetrapartite $\pi$-tangle of the GHZ state for bosonic and fermionic fields in Fig.\ref{Fig8} as a function of the Hawking temperature $T$, for mode frequencies $\omega=1, 2, 3$. For both fields, the entanglement decreases monotonically with increasing $T$, and higher‑frequency modes retain more entanglement. Furthermore, in the high‑temperature limit $T \to \infty$, the residual tetrapartite entanglement tends to a finite value independent of the mode frequency $\omega$. This behavior is consistent with the conclusions for the GHS dilaton black hole.

\begin{figure}
\begin{minipage}[t]{0.5\linewidth}
\centering
\includegraphics[width=3.5in,height=6.5cm]{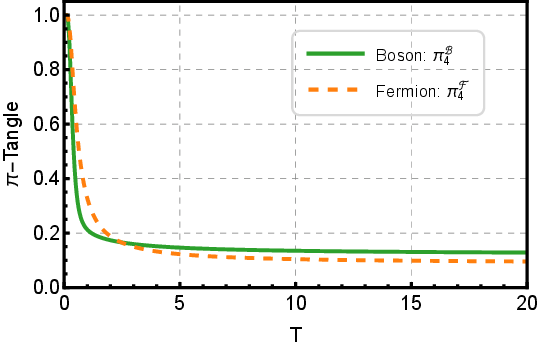}
\label{fig5}
\end{minipage}%
\caption{Tetrapartite $\pi$-tangle of bosonic and fermionic fields versus the Hawking temperature $T$ in Schwarzschild spacetime, with fixed $\omega=1$.}
\label{Fig9}
\end{figure}
Fig.\ref{Fig9} shows a direct comparison of bosonic and fermionic $\pi$-tangles as a function of $T$ for $\omega=1$ in the Schwarzschild background. A clear crossover occurs at $T\sim 2.2$: fermionic entanglement dominates at low temperatures, while bosonic entanglement becomes dominant in the high‑temperature regime. This confirms that the observed bosonic robustness is a universal feature rather than a model‑dependent artifact of the GHS extremal geometry. Since the Schwarzschild black hole reproduces the same qualitative crossover behavior, we conclude that this phenomenon is not a peculiarity of the GHS extremal geometry (i.e., not a model‑dependent artifact), but rather a universal feature arising from the fundamental difference in quantum statistics.

\section{N-partite genuine entanglement of GHZ state for fermionic field in dilaton spacetime}
Consider an N-partite GHZ state of a fermionic field shared among $N$ observers, where $m$ observers hover near the event horizon of a dilaton black hole and the remaining $q=N-m$ observers are located in the asymptotically flat region. Using the dilaton modes for fermionic field given in Eqs.(\ref{S22}) and (\ref{S23}), the state can be written as
\begin{eqnarray}\label{S44}
\left|GHZ\right\rangle_{N}^{\mathcal{F}}&&=\frac{1}{\sqrt{2}} \Bigg\{\left|0\right\rangle^{\otimes q}\Big[\frac{1}{\sqrt{e^{-8\pi\omega(M-\epsilon)}+1}}\left|0\right\rangle_{\mathrm{out}}\left|0\right\rangle_{\mathrm{in}}+\frac{1}{\sqrt{e^{8\pi\omega(M-\epsilon)}+1}}\left|1\right\rangle_{\mathrm{out}}\left|1\right\rangle_{\mathrm{in}}\Big]^{\otimes m}\notag\\
&&+\left|1\right\rangle^{\otimes q}\big(\left|1\right\rangle_{\mathrm{out}}\left|1\right\rangle_{\mathrm{in}}\big)^{\otimes m}\Bigg\}.
\end{eqnarray}
The corresponding density operator is
\begin{eqnarray}\label{S45}
\rho_{N}^{\mathcal{F}}&=&\frac{1}{2}\Bigg\{|0\rangle^{\otimes q}\Big[\frac{1}{\sqrt{e^{-8\pi\omega(M-\epsilon)}+1}}|0\rangle_{\mathrm{out}}|0\rangle_{\mathrm{in}}+\frac{1}{\sqrt{e^{8\pi\omega(M-\epsilon)}+1}}|1\rangle_{\mathrm{out}}|1\rangle_{\mathrm{in}}\Big]^{\otimes m}\langle0|^{\otimes q}\notag\\
&&\times\Big[\frac{1}{\sqrt{e^{-8\pi\omega(M-\epsilon)}+1}}\langle0|_{\mathrm{out}}\langle0|_{\mathrm{in}}+\frac{1}{\sqrt{e^{8\pi\omega(M-\epsilon)}+1}}\langle1|_{\mathrm{out}}\langle1|_{\mathrm{in}}\Big]^{\otimes m}\notag\\
&&+|0\rangle^{\otimes q}\Big[\frac{1}{\sqrt{e^{-8\pi\omega(M-\epsilon)}+1}}|0\rangle_{\mathrm{out}}|0\rangle_{\mathrm{in}}+\frac{1}{\sqrt{e^{8\pi\omega(M-\epsilon)}+1}}|1\rangle_{\mathrm{out}}|1\rangle_{\mathrm{in}}\Big]^{\otimes m}\notag\\
&&\times\langle1|^{\otimes q}\Big[\langle1|_{\mathrm{out}}\langle0|_{\mathrm{in}}\Big]^{\otimes m}+|1\rangle^{\otimes q}\Big[|1\rangle_{\mathrm{out}}|0\rangle_{\mathrm{in}}\Big]^{\otimes m}\langle 0|^{\otimes q}\Big[\frac{1}{\sqrt{e^{-8\pi\omega(M-\epsilon)}+1}}\langle0|_{\mathrm{out}}\langle0|_{\mathrm{in}} \notag\\
&&+\frac{1}{\sqrt{e^{8\pi\omega(M-\epsilon)}+1}}\langle1|_{\mathrm{out}}\langle1|_{\mathrm{in}}\Big]^{\otimes m}+|1\rangle^{\otimes q}\Big[|1\rangle_{\mathrm{out}}|0\rangle_{\mathrm{in}}\Big]^{\otimes m}\langle1|^{\otimes q}\Big[\langle1|_{\mathrm{out}}\langle0|_{\mathrm{in}}\Big]^{\otimes m}\Bigg\}.
\end{eqnarray}
By tracing out the physically inaccessible modes, the corresponding reduced density operator is given by
\begin{eqnarray}\label{S415}
(\rho_{N}^{\mathcal{F}})^{\operatorname{Tr}_{II}} &=& \frac{1}{2} \Bigg\{ \big |0\rangle^{\otimes q} \langle 0|^{\otimes q} \otimes \Big[\big[\frac{1}{\sqrt{e^{-8\pi\omega(M-\epsilon)}+1}}\big]^{2} |0\rangle_{\mathrm{out}} \langle 0|_{\mathrm{out}} + \big[\frac{1}{\sqrt{e^{8\pi\omega(M-\epsilon)}+1}}\big]^{2} |1\rangle_{\mathrm{out}} \langle 1|_{\mathrm{out}} \Big]^{\otimes m} \notag \\
& +& |0\rangle^{\otimes q} \langle 1|^{\otimes q} \otimes \big[\frac{1}{\sqrt{e^{-8\pi\omega(M-\epsilon)}+1}} |0\rangle_{\mathrm{out}} \langle 1|_{\mathrm{out}} \big]^{\otimes m} \notag \\
& +&|1\rangle^{\otimes q} \langle 0|^{\otimes q} \otimes \big[\frac{1}{\sqrt{e^{-8\pi\omega(M-\epsilon)}+1}} |1\rangle_{\mathrm{out}} \langle 0|_{\mathrm{out}} \big]^{\otimes m} \notag \\
& +& \big |1\rangle^{\otimes q} \langle 1|^{\otimes q} \otimes \big( |1\rangle_{\mathrm{out}} \langle 1|_{\mathrm{out}} \big)^{\otimes m} \Bigg\}.
\end{eqnarray}

We now focus on a tetrapartite GHZ state shared by four observers, Alice ($A$), Bob ($B$), Charlie ($C$), and David ($D$), with the latter three observers ($m=3$) situated in the vicinity of the event horizon. Due to the causal disconnection between the interior and exterior regions of the black hole, tracing over the inaccessible modes inside the event horizon leads to the reduced density matrix
\begin{eqnarray}\label{S46}
\rho_{ABCD}^{\mathcal{F}}&=&\frac{1}{2} \Bigg\{\big[\frac{1}{\sqrt{e^{-8\pi\omega(M-\epsilon)}+1}}\big] ^{6}|0000\rangle\langle0000|+\big[\frac{1}{\sqrt{e^{-8\pi\omega(M-\epsilon)}+1}}\big]^{4}\big[\frac{1}{\sqrt{e^{8\pi\omega(M-\epsilon)}+1}}\big]^{2}\notag\\
&&\times|0001\rangle\langle0001|+\big[\frac{1}{\sqrt{e^{-8\pi\omega(M-\epsilon)}+1}}\big]^{4}\big[\frac{1}{\sqrt{e^{8\pi\omega(M-\epsilon)}+1}}\big]^{2}|0010\rangle\langle0010|\notag\\
&&+\big[\frac{1}{\sqrt{e^{-8\pi\omega(M-\epsilon)}+1}}\big]^{4}\big[\frac{1}{\sqrt{e^{8\pi\omega(M-\epsilon)}+1}}\big]^{2}|0100\rangle\langle0100|\notag\\
&&+\big[\frac{1}{\sqrt{e^{-8\pi\omega(M-\epsilon)}+1}}\big]^{2}\big[\frac{1}{\sqrt{e^{8\pi\omega(M-\epsilon)}+1}}\big]^{4}|0011\rangle\langle0011|\notag\\
&&+\big[\frac{1}{\sqrt{e^{-8\pi\omega(M-\epsilon)}+1}}\big]^{2}\big[\frac{1}{\sqrt{e^{8\pi\omega(M-\epsilon)}+1}}\big]^{4}|0101\rangle\langle0101|\notag\\
&&+\big[
\frac{1}{\sqrt{e^{-8\pi\omega(M-\epsilon)}+1}}\big]^{3}|0000\rangle\langle1111|+\big[\frac{1}{\sqrt{e^{8\pi\omega(M-\epsilon)}+1}}\big] ^{6}|0111\rangle\langle0111|\notag\\
&&+\big[\frac{1}{\sqrt{e^{-8\pi\omega(M-\epsilon)}+1}}\big]^{2}\big[\frac{1}{\sqrt{e^{8\pi\omega(M-\epsilon)}+1}}\big]^{4}|0110\rangle\langle0110|+|1111\rangle\langle1111| \Bigg\}.
\end{eqnarray}
Applying the partial-transposition rules given in Eq.(\ref{S37}) together with the definition of negativity in Eq.(\ref{S28}), we obtain the one-tangle between Alice and the remaining subsystem as
\begin{eqnarray}\label{S47}
N_{A(BCD)}^{\mathcal{F}}=\sqrt{\frac{\big[\frac{1}{\sqrt{e^{8\pi\omega(M-\epsilon)}+1}}\big]^{12}}{4}+\big[\frac{1}{\sqrt{e^{-8\pi\omega(M-\epsilon)}+1}}\big]^{6}}-\frac{\big[\frac{1}{\sqrt{e^{8\pi\omega(M-\epsilon)}+1}}\big]^{6}}{2}.
\end{eqnarray}
Furthermore, due to the symmetry among Bob, Charlie, and David, the corresponding one-tangles are identical and read
\begin{eqnarray}\label{S48}	
N_{B(ACD)}^{\mathcal{F}}=N_{C(ABD)}^{\mathcal{F}}=N_{D(ABC)}^{\mathcal{F}}&&=\sqrt{\frac{\big[\frac{1}{\sqrt{e^{-8\pi\omega(M-\epsilon)}+1}}\big]^{8}\big[\frac{1}{\sqrt{e^{8\pi\omega(M-\epsilon)}+1}}\big]^{4}}{4}+\big[\frac{1}{\sqrt{e^{-8\pi\omega(M-\epsilon)}+1}}\big]^{6}}\notag\\
&&-\frac{\big[\frac{1}{\sqrt{e^{-8\pi\omega(M-\epsilon)}+1}}\big]^{4}\big[\frac{1}{\sqrt{e^{8\pi\omega(M-\epsilon)}+1}}\big]^{2}}{2}.
\end{eqnarray}

To quantify the entanglement accessible to observers in the asymptotically flat region, we evaluate the partial transpose of the reduced density matrix with respect to one of their qubits. Due to permutation symmetry, the resulting spectrum is independent of the specific choice. Taking the partial transpose with respect to the first qubit, we obtain
\begin{align}
\label{S149}
\big(\rho_{N}^{\mathcal{F}}\big)^{T_{1}} = &\frac{1}{2} \Bigg\{ \big  |0\rangle^{\otimes q} \langle 0|^{\otimes q} \otimes \Big[ \big[\frac{1}{\sqrt{e^{-8\pi\omega(M-\epsilon)}+1}}\big]^{2} |0\rangle_{\mathrm{out}} \langle 0|_{\mathrm{out}} + \big[\frac{1}{\sqrt{e^{8\pi\omega(M-\epsilon)}+1}}\big]^{2} |1\rangle_{\mathrm{out}} \langle 1|_{\mathrm{out}} \Big]^{\otimes m} \notag\\
& + |1\rangle |0\rangle^{\otimes (q-1)} \langle 0| \langle 1|^{\otimes (q-1)} \otimes \big[ \frac{1}{\sqrt{e^{-8\pi\omega(M-\epsilon)}+1}} |0\rangle_{\mathrm{out}} \langle 1|_{\mathrm{out}} \big]^{\otimes m} \notag \\
& + |0\rangle |1\rangle^{\otimes (q-1)} \langle 1| \langle 0|^{\otimes (q-1)} \otimes \big[\frac{1}{\sqrt{e^{-8\pi\omega(M-\epsilon)}+1}} |1\rangle_{\mathrm{out}} \langle 0|_{\mathrm{out}} \big]^{\otimes m} \notag \\
& + |1\rangle^{\otimes q} \langle 1|^{\otimes q} \otimes \big( |1\rangle_{\mathrm{out}} \langle 1|_{\mathrm{out}} \big)^{\otimes m} \Bigg\}.
\end{align}
The negative eigenvalue of $(\rho_{N}^{\mathcal{F}})^{T_{1}}$ is found to be
\begin{eqnarray}\label{S49}
\lambda_{-}=-\frac{1}{2}\Bigg\{\sqrt{\frac{\big[\frac{1}{\sqrt{e^{8\pi\omega(M-\epsilon)}+1}}\big]^{4m}}{4}+\big[\frac{1}{\sqrt{e^{-8\pi\omega(M-\epsilon)}+1}}\big]^{2m}}-\frac{\big[\frac{1}{\sqrt{e^{8\pi\omega(M-\epsilon)}+1}}\big]^{2m}}{2}\Bigg\}.
\end{eqnarray}
Substituting this into Eq.(\ref{S28}) yields the negativity for the observer
\begin{eqnarray}\label{S50}
N_{\mathrm{flat}}^{\mathcal{F}}=\sqrt{\frac{\big[\frac{1}{\sqrt{e^{8\pi\omega(M-\epsilon)}+1}}\big]^{4m}}{4}+\big[\frac{1}{\sqrt{e^{-8\pi\omega(M-\epsilon)}+1}}\big]^{2m}}-\frac{\big[\frac{1}{\sqrt{e^{8\pi\omega(M-\epsilon)}+1}}\big]^{2m}}{2},
\end{eqnarray}
Eq.(\ref{S50}) indicates that quantum entanglement perceived by the observers in the asymptotically flat region decreases with increasing number of near-horizon qubits $m$.
Finally, starting from Eq.(\ref{S48}), an analysis analogous to that for the bosonic field can be performed for observers near the event horizon. The entanglement measured by an arbitrary one of the $m$ near-horizon observers is given by
\begin{eqnarray}\label{S53}	
N_{\mathrm{horizon}}^{\mathcal{F}}&&=\sqrt{\frac{\big[\frac{1}{\sqrt{e^{-8\pi\omega(M-\epsilon)}+1}}\big]^{4(m-1)}\big[\frac{1}{\sqrt{e^{8\pi\omega(M-\epsilon)}+1}} \big]^{4}}{4}+\big[\frac{1}{\sqrt{e^{-8\pi\omega(M-\epsilon)}+1}}\big]^{2m}}\notag\\
&&-\frac{\big[\frac{1}{\sqrt{e^{-8\pi\omega(M-\epsilon)}+1}}\big]^{2(m-1)}\big[\frac{1}{\sqrt{e^{8\pi\omega(M-\epsilon)}+1}}\big]^{2}}{2}.
\end{eqnarray}


\begin{thebibliography}{99}
\bibitem{LW1}
H. J. Kimble, The quantum internet, Nature {\bf453}, 1023 (2008).
\bibitem{LW2}
W. D\"{u}r, R. Lamprecht, and S. Heusler, Towards a quantum internet, Eur. J. Phys. {\bf38}, 043001 (2017).
\bibitem{LW3}
C. Simon, Towards a global quantum network, Nat. Photonics {\bf11}, 678 (2017).
\bibitem{LW4}
S. Wehner, D. Elkouss, and R. Hanson, Quantum internet: A vision for the road ahead, Science {\bf362}, 6412 (2018).
\bibitem{LW5}
A. S. Cacciapuoti, M. Caleffi, F. Tafuri, F. S. Cataliotti, S. Gherardini, and G. Bianchi, Quantum Internet: Networking Challenges in Distributed Quantum Computing, IEEE Network {\bf34}, 137 (2020).
\bibitem{LW6}
M. Navascu\'{e}s, E. Wolfe, D. Rosset, and A. Pozas-Kerstjens, Genuine Network Multipartite Entanglement, Phys. Rev. Lett. {\bf125}, 240505 (2020).
\bibitem{LW7}
M. Navascu\'{e}s and E. Wolfe, The Inflation Technique Completely Solves the Causal Compatibility Problem, J. Causal Inference {\bf8}, 70 (2020).
\bibitem{LW8}
T. Kraft, S. Designolle, C. Ritz, N. Brunner, O. G\"{u}hne, and M. Huber, Quantum entanglement in the triangle network, Phys. Rev. A {\bf103}, L060401 (2021).
\bibitem{LW9}
E. Wolfe, A. Pozas-Kerstjens, M. Grinberg, D. Rosset, A. Ac\'{i}n, and M. Navascu\'{e}s, Quantum Inflation: A General Approach to Quantum Causal Compatibility, Phys. Rev. X {\bf11}, 021043 (2021).
\bibitem{LW10}
T. Kraft, C. Spee, X. D. Yu, and O. G\"{u}hne, Characterizing quantum networks: Insights from coherence theory, Phys. Rev. A {\bf103}, 052405 (2021).
\bibitem{LW11}
K. Hansenne, Z. P. Xu, T. Kraft, and O. G\"{u}hne, Symmetries in quantum networks lead to no-go theorems for entanglement distribution and to verification techniques, Nat. Commun. {\bf13}, 496 (2022).
\bibitem{LW12}
N. K. H. Li, X. Dai, M. H. Mu\~{n}oz-Arias, K. Reuer, M. Huber, and N. Friis, Detecting genuine multipartite entanglement in multi-qubit devices with restricted measurements, Nat. Commun. {\bf17}, 1707 (2026).
\bibitem{LW13}
P. Contreras-Tejada, C. Palazuelos, and J. I. d. Vicente, Genuine Multipartite Nonlocality Is Intrinsic to Quantum Networks, Phys. Rev. Lett. {\bf126}, 040501 (2021).
\bibitem{LW14}
P. Contreras-Tejada, C. Palazuelos, and J. I. d. Vicente, Asymptotic Survival of Genuine Multipartite Entanglement in Noisy Quantum Networks Depends on the Topology, Phys. Rev. Lett. {\bf128}, 220501 (2022).
\bibitem{LW15}
S. Morelli, D. Sauerwein, M. Skotiniotis, and N. Friis, Metrology-assisted entanglement distribution in noisy quantum networks, Quantum {\bf6}, 722 (2022).
\bibitem{LW16}
J. C. Besse, K. Reuer, M. C. Collodo, A. Wulff, L. Wernli, A. Copetudo, D. Malz, P. Magnard, A. Akin, M. Gabureac, G. J. Norris, J. I. Cirac, A. Wallraff, and C. Eichler, Realizing a deterministic source of multipartite-entangled photonic qubits, Nat. Commun. {\bf11}, 4877 (2020).
\bibitem{LW17}
M. Pompili, S. L. N. Hermans, S. Baier, H. K. C. Beukers, P. C. Humphreys, R. N. Schouten, R. F. L. Vermeulen, M. J. Tiggelman, L. d. S. Martins, B. Dirkse, S. Wehner, and R. Hanson, Realization of a multinode quantum network of remote solid-state qubits, Science {\bf372}, 259 (2021).
\bibitem{LW18}
A. Ruskuc, C. J. Wu, E. Green, S. L. N. Hermans, W. Pajak, J. Choi, and A. Faraon, Multiplexed entanglement of multiemitter quantum network nodes, Nature {\bf639}, 54 (2025).
\bibitem{LW19}
J. Shi, S. Zhang, Y. Wu, Y. Sun, Y. Liang, H. Wang, Y. Pu, and L. Duan, Scalable and Modular Generation of W-State Entanglements via Memory-Enhanced Fusion, Phys. Rev. Lett. {\bf135}, 150802 (2025).
\bibitem{LW20}
M. Canteri, J. Bate, I. Mishra, N. Friis, V. Krutyanskiy, and B. P. Lanyon, Generation of multipartite photonic entanglement using a trapped-ion quantum processing node, arXiv:2510.15693 (2025).
\bibitem{LW21}
R. A. Bertlmann and N. Friis, Modern Quantum Theory: From Quantum Mechanics to Entanglement and Quantum Information (Oxford University Press, Oxford, U.K., 2023).
\bibitem{LW22}
G. T\'{o}th, Multipartite entanglement and high-precision metrology, Phys. Rev. A {\bf85}, 022322 (2012).
\bibitem{LW23}
R. Raussendorf and H.J. Briegel, A One-Way Quantum Computer, Phys. Rev. Lett. {\bf86}, 5188 (2001).
\bibitem{LW24}
H. J. Briegel and R. Raussendorf, Persistent Entanglement in Arrays of Interacting Particles, Phys. Rev. Lett. {\bf86}, 910 (2001).
\bibitem{LW25}
A. J. Scott, Multipartite entanglement, quantum-error-correcting codes, and entangling power of quantum evolutions, Phys. Rev. A {\bf69}, 052330 (2004).
\bibitem{LW26}
M. Epping, H. Kampermann, C. Macchiavello, and D. Bruß, Multi-partite entanglement can speed up quantum key distribution in networks, New J. Phys. {\bf19}, 093012 (2017).
\bibitem{LW27}
M. Pivoluska, M. Huber, and M. Malik, Layered quantum key distribution, Phys. Rev. A {\bf97}, 032312 (2018).
\bibitem{LW28}
J. Ribeiro, G. Murta, and S. Wehner, Fully device-independent conference key agreement, Phys. Rev. A {\bf97}, 022307 (2018).
\bibitem{LW29}
S. B\"{a}uml and K. Azuma, Fundamental limitation on quantum broadcast networks, Quantum Sci. Technol. {\bf2}, 024004 (2017).
\bibitem{LW30}
H. Yamasaki, A. Pirker, M. Murao, W. D\"{u}r, and B. Kraus, Multipartite entanglement outperforming bipartite entanglement under limited quantum system sizes, Phys. Rev. A {\bf98}, 052313 (2018).
\bibitem{SDF1}
W. Liu, C. Wen, J. Wang, Lorentz violation alleviates gravitationally induced entanglement degradation, J. High Energy Phys. {\bf2025}, 184 (2025).

\bibitem{SDF2}
S. Sen, A. Mukherjee and S. Gangopadhyay, Entanglement degradation as a tool to detect signatures of modified gravity, Phys. Rev. D {\bf109}, 046012 (2024).

\bibitem{SDF3}
E. Mart\'{\i}n-Mart\'{\i}nez, L. J. Garay and J. Le\'{o}n, Unveiling quantum entanglement degradation near a Schwarzschild black hole,  Phys. Rev. D {\bf82}, 064006 (2010).


\bibitem{SDF4}
Q. Pan and J. Jing, Hawking radiation, entanglement, and teleportation in the background of an asymptotically flat static black hole, Phys. Rev. D {\bf78}, 065015 (2008).

\bibitem{SDF5}
H. E. Camblong, A. Chakraborty, P. Lopez-Duque, C. R. Ord\'{o}\~{n}ez, Entanglement degradation in causal diamonds,
Phys. Rev. D {\bf109}, 105003 (2024).

\bibitem{SDF6}
A. Ali, S. Al-Kuwari, M. Ghominejad, M. T. Rahim, D. Wang and S. Haddadi, Quantum characteristics near event horizons, Phys. Rev. D {\bf110}, 064001 (2024).


\bibitem{SDF7}
I. Fuentes-Schuller and R. B. Mann, Alice Falls into a Black Hole: Entanglement in Noninertial Frames, Phys. Rev. Lett. {\bf95}, 120404 (2005).

\bibitem{SDF8}
P. M. Alsing, I. Fuentes-Schuller, R. B. Mann and T. E. Tessier, Entanglement of Dirac fields in noninertial frames, Phys. Rev. A {\bf74}, 032326 (2006).


\bibitem{SDF9}
S. M. Wu, X. W. Fan, R. D. Wang, H. Y. Wu, X. L. Huang and H. S. Zeng, Does Hawking effect always degrade fidelity of quantum teleportation in Schwarzschild spacetime?, J. High Energy Phys. {\bf2023}, 232 (2023).

\bibitem{SDF10}
M. M. Du, H. W. Li, S. T. Shen, X. J. Yan, X. Y. Li, L. Zhou, W. Zhong and Y. B. Sheng, Maximal steered coherence in the background of Schwarzschild space-time, Eur. Phys. J. C {\bf84}, 450 (2024).

\bibitem{SDF11}
S. Elghaayda, A. Ali, S. Al-Kuwari and M. Mansour, Physically accessible and inaccessible quantum correlations of Dirac fields in Schwarzschild spacetime, Phys. Lett. A {\bf525}, 129915 (2024).


\bibitem{SDF12}
J. K. Basak, D. Giataganas, S. Mondal and W. Y. Wen, Reflected entropy and Markov gap in noninertial frames, Phys. Rev. D {\bf108}, 125009 (2023).


\bibitem{SDF13}
W. M. Li, S. M. Wu, Bosonic and fermionic coherence of N-partite states in the background of a dilaton black hole,  J. High Energy Phys. {\bf2024}, 144 (2024).




\bibitem{SDF14}
C. y. Liu, Z. w. Long and Q. l. He, Quantum coherence and quantum Fisher information of Dirac particles in curved spacetime under decoherence, Phys. Lett. B {\bf857}, 138991 (2024).

\bibitem{SDF15}
T. Zhang, X. Wang and S. M. Fei, Hawking effect can generate physically inaccessible genuine tripartite nonlocality, Eur. Phys. J. C {\bf83}, 607 (2023).

\bibitem{SDF16}
S. M. Wu, H. Y. Wu, Y. X. Wang, J. Wang, Gaussian tripartite steering in Schwarzschild black hole, Phys. Lett. B {\bf865},  139493 (2025).

\bibitem{SDF17}
G. Adesso, I. Fuentes-Schuller and M. Ericsson, Continuous variable entanglement sharing in non-inertial frames, Phys. Rev. A {\bf76}, 062112 (2007).

\bibitem{SDF18}
S. M.  Wu, X. W. Teng, W. M. Li, Y. X. Wang, J. Lu, Nonseparability of multipartite systems in dilaton black hole, JCAP {\bf09}, 030  (2025).


\bibitem{SDF19}
G. W. Mi, X. Huang, S. M. Fei, T. Zhang, Quantumness near the Schwarzschild black hole based on W-state,
Ann. Phys. (Berlin)  {\bf10}, 1002 (2025).

\bibitem{SDF20}
S. Harikrishnan, S. Jambulingam, P. P. Rohde and C. Radhakrishnan, Accessible and inaccessible quantum coherence in relativistic quantum systems, Phys. Rev. A {\bf105}, 052403 (2022).

\bibitem{SDF21}
G. W. Mi, X. Huang, S. M. Fei, T. Zhang, Genuine four-partite Bell nonlocality in the curved spacetime, Eur. Phys. J. C {\bf85}, 354, (2025).


\bibitem{SDF22}
T. Y. Wang and D. Wang, Entropic uncertainty relations in Schwarzschild space-time, Phys. Lett. B {\bf855}, 138876 (2024).


\bibitem{SDF23}
S. Bellucci, V. Kh. Kotanjyan, A. A. Saharian, Fermionic condensate and the mean energy-momentum tensor
in the Fulling-Rindler vacuum, Phys. Rev. D {\bf108}, 085014 (2023).

\bibitem{SDF24}
G. W. Mi, X. Huang, S. M. Fei, T. Zhang, Impact of the Hawking Effect on the fully entangled
fraction of three-qubit states in Schwarzschild spacetime, Ann. Phys. (Berlin) {\bf537}, 2400308
(2024).

\bibitem{SDF25}
H. Wu and L. Chen, Orbital angular momentum entanglement in noninertial reference frame, Phys. Rev. D {\bf107}, 065006 (2023).


\bibitem{SDF26}
L. J. Li, F. Ming, X. K. Song, L. Ye and D. Wang, Quantumness and entropic uncertainty in curved space-time, Eur. Phys. J. C  {\bf82}, 726 (2022).

\bibitem{SDF27}
A. Belfiglio, O. Luongo, S. Mancini, Quantum entanglement in cosmology, Phys. Rep. {\bf1146}, 1-47 (2025).


\bibitem{SDF28}
X. Liu, W. Liu, Z. Liu, J. Wang, Harvesting correlations from BTZ black hole coupled to a Lorentz-violating vector field, J. High Energy Phys. {\bf2025}, 94 (2025).


\bibitem{SDF29}
Z. D. Wei, W. Han, Y. J. Zhang, Z. X. Man, Y. J. Xia, H. Fan, Effect of the gravitational redshift on the precision of phase estimation, Phys. Rev. D \textbf{111},  026007 (2025).


\bibitem{SDF30}
W. Izquierdo,  J. Beltran,  E. Arias, Enhancement of harvesting vacuum entanglement in Cosmic String Spacetime, J. High Energy Phys. {\bf2025}, 049 (2025).


\bibitem{SDF31}
Z. Liu, W. Liu, X. Liu, J. Wang, Wormhole-Induced correlation: A Link Between Two Universes, arXiv:2510.04005.

\bibitem{SDF32}
S. M. Wu, C. X. Wang, D. D. Liu, X. L. Huang, H. S. Zeng, Would quantum coherence be increased by curvature effect in de Sitter space?, J. High Energy Phys. {\bf2023}, 115  (2023).

\bibitem{SDF33}
A. Chakraborty,  L. Hackl,  M. Zych, Entanglement harvesting in quantum superposed spacetime, Phys. Rev. D {\bf111}, 104052 (2025).

\bibitem{SDF34}
Z. Liu, R. Q. Yang, H. Fan, J. Wang, Simulation of the massless Dirac field in 1+1D curved spacetime, Phys. Mech. Astron. {\bf68}, 290411 (2025).

\bibitem{SDF35}
S. M. Wu, R. D. Wang, X. L. Huang, Z. Wang, Does gravitational wave assist vacuum steering and Bell nonlocality?, J. High Energy Phys. {\bf2024}, 155 (2024).

\bibitem{SDF36}
Z. Liu, Y. Li, Z. Tian, J. Wang, Scrambling-Enhanced Quantum Battery Charging in Black Hole Analogues, Adv. Sci.  e20281 (2025).


\bibitem{SDF37}
X. Liu, Z. Tian, J. Jing, Entanglement dynamics in $\kappa$-deformed
spacetime, Sci. China Phys. Mech. Astron. {\bf67}, 100411 (2024).


\bibitem{SDF38}
S. Barman, I. Chakraborty, S. Mukherjee, Signatures of gravitational wave memory in the radiative process of entangled quantum probes, Phys. Rev. D {\bf111}, 025021 (2025).


\bibitem{SDF39}
J. Foo,  R. B. Mann, M. Zych, Entanglement amplification between superposed detectors in flat and curved spacetimes, 
Phys. Rev. D {\bf103}, 065013  (2021).


\bibitem{SDF40}
X. Liu, Z. Tian, J. Jing, Dissipation suppression for an Unruh-DeWitt battery with a reflecting boundary. Sci. China Phys. Mech.
Astron. {\bf68}, 100412 (2025).

\bibitem{SDF41}
X. Liu, Z. Tian, J. Wang, J. Jing, Protecting quantum coherence of two-level atoms from vacuum fluctuations of electromagnetic field, Ann. Phys.  {\bf366}, 102 (2016).

\bibitem{SDF42}
J. Foo, C. S. Arabaci, M. Zych, R. B. Mann, Quantum Signatures of Black Hole Mass Superpositions, Phys. Rev. Lett. {\bf129}, 181301  (2022).

\bibitem{SDF43}
J. Foo, C. S. Arabaci, M. Zych,  R. B. Mann, Quantum superpositions of Minkowski spacetime, Phys. Rev. D {\bf107}, 045014 (2023).


\bibitem{SDF44}
M. M. Du, H. W. Li, Z. Tao, S. T. Shen, X. J. Yan, X. Y. Li, W. Zhong, Y. B. Sheng and L. Zhou, Basis-independent quantum coherence and its distribution under relativistic motion, Eur. Phys. J. C {\bf84}, 838 (2024).

\bibitem{SDF45}
Y. Ji, J. Zhang and H. Yu, Entanglement harvesting in cosmic string spacetime, J. High Energy Phys. {\bf2024}, 161 (2024).

\bibitem{SDF46}
Z. Liu, J. Zhang and H. Yu, Harvesting correlations from vacuum quantum fields in the presence of a refecting boundary, J. High Energy Phys. {\bf2023}, 184 (2023).

\bibitem{SDF47}
Q. Liu, T. Liu, C. Wen, J. Wang, Optimal quantum strategy for locating Unruh channels, Phys. Rev. A {\bf110}, 022428 (2024).

\bibitem{SDF48}
Y. K. Zhang, L. J. Li, X. K. Song, L. Ye and D. Wang, Entropic uncertainty and quantum non-classicality of Unruh-Dewitt detectors in relativity, Phys. Lett. B {\bf858}, 139063 (2024).

\bibitem{SDF49}
S. H. Li, S. H. Shang, S. M. Wu, Does acceleration always degrade quantum entanglement for tetrapartite Unruh-DeWitt detectors?, J. High Energy Phys. {\bf2025}, 214  (2025).


\bibitem{SDF50}
T. Gonzalez-Raya, S. Pirandola and M. Sanz, Satellite-based entanglement distribution and quantum teleportation with continuous variables, Commun. Phys. {\bf7},  126 (2024).


\bibitem{SDF51}
Y. Tang, W. Liu, Z. Liu, J. Wang, Can the latent signatures of quantum superposition be detected
through correlation harvesting?, J. High Energy Phys. {\bf2026}, 045 (2026). 

\bibitem{SDF52}
Y. Tang, W. Liu, J. Wang, Observational signature of Lorentz violation in acceleration radiation, Eur.
Phys. J. C {\bf 85}, 1108 (2025).


\bibitem{SDF53}
X. Liu, C. Wen, J. Wang, Quantum coherence of continuous variables in the black hole quantum atmosphere, Phys. Lett. B {\bf873}, 140185 (2026).


\bibitem{SDF54}
I. Agullo, A. Delhom, \'A. Parra-L\'opez, Toward the observation of entangled pairs in BEC analog expanding universes, Phys. Rev. D \textbf{110}, 125023 (2024).

\bibitem{SDF55}
S. M. Wu and H. S. Zeng, Genuine tripartite nonlocality and entanglement in curved spacetime, Eur. Phys. J. C {\bf82}, 4 (2022).

\bibitem{SDF56}
A. A. Svidzinsky, M. O. Scully,  W. Unruh, Minkowski vacuum entanglement and accelerated oscillator chains, Phys. Rev. D {\bf111}, 045022 (2025).

\bibitem{SDF57}
N. Arya, M. Zych, Selective Amplification of a Gravitational Wave Signal Using an Atomic Array, 
arXiv:2408.12436. 


\bibitem{AGL1}
J. Paczos, N. Arya, S. Qvarfort, D. Braun, M. Zych, Gravitational Wave Imprints on Spontaneous Emission, Phys. Rev. Lett. {\bf136}, 113201 (2026).

\bibitem{AGL2}
S. C. Liu, L. H. Liu, B. Li, H. Q. Zhang, P. Z. He, A quantum information method for early universe with non-trivial sound speed, Fortschr. Phys. {\bf74}, e70081 (2026).

\bibitem{AGL3}
A. Deswal, N. Arya, K. Lochan, S. K. Goyal, Time-resolved and Superradiantly Amplified Unruh Effect,
Phys. Rev. Lett. {\bf135}, 183601 (2025).

\bibitem{AGL99}
A. Mukherjee, S. Gangopadhyay, P. H. M. Barros, H. A. S. Costa, Transition rates and their applications in accelerated single-qubit for fermionic spinor field coupling, arXiv:2512.09144


\bibitem{AGL100}
A. Mukherjee, S. Gangopadhyay, A. S. Majumdar, Fulling-Davies-Unruh effect for accelerated two-level single and entangled atomic systems, Phys. Rev. D {\bf108}, 085018 (2023).


\bibitem{AGL101}
Q. Xiao, Y. Chen, T. Liu, Effects of Lorentz symmetry breaking on quantum coherence in an expanding universe, 
 Phys. Lett. B {\bf872}, 140133  (2026).

\bibitem{AGL102}
P. K. Kumawat, S. Barman, B. R. Majhi, Equivalence in virtual transitions between uniformly accelerated and static atoms: from a bird's eye,  JCAP {\bf02}, 046  (2025).

\bibitem{AGL103}
G. Ciliberto, S. Emig, N. Pavloff, M. Isoard, Violation of Bell inequalities in an analog black hole, Phys. Rev. A {\bf109}, 063325 (2024).

\bibitem{AGL104}
T. Li, L. H. Liu, Inflationary Krylov complexity, J. High Energy Phys. {\bf2024}, 123 (2024).








\bibitem{J9}
G.W. Gibbons, Antigravitating black hole solitons with scalar hair in N=4 supergravity, Nucl. Phys. B {\bf207}, 337 (1982).

\bibitem{J10}
A. Gare\'{\i}a, D. Galtsov, and O. Kechkin, Class of stationary axisymmetric solutions of the Einstein-Maxwell-Dilaton-Axion field equations, Phys. Rev. Lett. {\bf74}, 1276 (1995).

\bibitem{J11}
G.W. Gibbons and K. Maeda, Black holes and membranes in higher-dimensional theories with dilaton fields, Nucl. Phys. B {\bf298}, 741 (1988).














\bibitem{A1}
J. Wang, Q. Pan, S. Chen, and J. Jing, Entanglement of coupled massive scalar field in background of dilaton black hole, Phys. Lett. B {\bf677}, 186 (2009).



 \bibitem{A3}
D. Garfinkle, G. T. Horowitz, and A. Strominger, Charged black holes in string theory, Phys. Rev. D {\bf45}, 3888 (1992).


\bibitem{A4}
Q. Xiao, C. Wen, J. Jing, and J. Wang, Generation of quantum coherence for continuous variables between causally disconnected regions in dilaton spacetime, Eur. Phys. J. C {\bf82}, 893 (2022).





\bibitem{B2}
C. Liu, Z. Long, Q. He,  Would the fidelity of quantum teleportation be increased by a local filtering operation near a dilaton black hole under decoherence?, Eur. Phys. J. C {\bf85}, 926 (2025).

\bibitem{B3}
F. Shahbazi, S. Haseli, H. Dolatkha, and S. Salimi, Entropic uncertainty relation in Garfinkle-Horowitz-Strominger dilation black hole, JCAP {\bf10}, 047 (2020).


\bibitem{B4}
S. M. Wu, J. X. Li, X. Y. Jiang, X. W. Teng, X. L. Huang, and J.  Lu, Fermionic steering is not nonlocal in the background of dilaton black hole, Eur. Phys. J. C {\bf84}, 161 (2024).

\bibitem{B5}
Y. C. Ou and H. Fan, Monogamy inequality in terms of negativity for three-qubit states, Phys. Rev. A {\bf75}, 062308 (2007).

\bibitem{B9}
R. Horodecki, P. Horodecki, M. Horodecki, and K. Horodecki, Quantum entanglement, Rev. Mod. Phys. {\bf81}, 865 (2009).

\bibitem{B6}
W. B. Xing, X. M. Hu, Y. Guo, B. H. Liu, C. F. Li, and G. C. Guo, Preparation of multiphoton high-dimensional GHZ states, Opt. Express {\bf31}, 24887 (2023).

\bibitem{B7}
Y. M. Liu, Z. Y. Wang , J. Liu and Z. J. Zhang, Remote Preparation of Three-Particle GHZ Class States, Commun. Theor. Phys. {\bf49}, 359 (2008).

\bibitem{B8}
N. Li, X. Lu, and J. M. Liu, Experimental realization of deterministic joint remote preparation of an arbitrary two-qubit pure state via GHZ states, Sci. China Inf. Sci. {\bf67}, 139402 (2024).

\bibitem{B10}
S. Chandrasekhar, The Mathematical Theory of Black Holes, Oxford University Press, (1983).
\bibitem{B11}
R. Casadio, A. Fabbri, and L. Mazzacurati, New black holes in the brane world?, Phys. Rev. D {\bf65}, 084040 (2002).
\bibitem{B12}
W. G. Unruh, Notes on black-hole evaporation, Phys. Rev. D {\bf14}, 870 (1976).
\bibitem{B13}
T. Damour and R. Ruffini, Black-hole evaporation in the Klein-Sauter-Heisenberg-Euler formalism, Phys. Rev. D {\bf14}, 332 (1976).
\bibitem{B14}
S. Sannan, Heuristic derivation of the probability distributions of particles emitted by a black hole, Gen. Relativ. Gravit. {\bf20}, 239 (1988).
\bibitem{B15}
Z. Zhao and Y. X. Gui, The connection between Unruh scheme and Damour-Ruffini scheme in Rindler space-time and eta-5 space-time, Nuovo Cimento Soc. Ital. Fis. B {\bf109}, 355 (1994).
\bibitem{B16}
N. D. Birrell and P. C. W. Davies, Quantum Fields in Curved Space, Cambridge University Press, Cambridge, England, (1982).


\bibitem{B18}
D. Ahn, Final state boundary condition of the Schwarzschild black hole, Phys. Rev. D {\bf74}, 084010 (2006).
\bibitem{B17}
S. M. Barnett and P. M. Radmore, Methods in Theoretical
Quantum Optics, Oxford University Press, New York, {\bf67}, (1997).

\bibitem{B19}
R. Kerner and R. B. Mann, Tunnelling, temperature, and Taub-NUT black holes, Phys. Rev. D {\bf73}, 104010 (2006).

\bibitem{B20}
Q. Q. Jiang, S. Q. Wu, and X. Cai, Hawking radiation from dilatonic black holes via anomalies, Phys. Rev. D {\bf75}, 064029 (2007).

\end{thebibliography}
\end{document}